\pdfoutput=1

\documentclass[a4paper,11pt]{article}
\usepackage{jheppub}

\usepackage{bm}

\usepackage{color}
\usepackage[usenames,dvipsnames,svgnames,table]{xcolor}

\usepackage{amsmath}
\usepackage{amssymb}
\usepackage{graphicx}
\usepackage{slashed}
\usepackage{soul}
\usepackage{multirow}
\usepackage{subcaption}

\newcommand{\MET}{\ensuremath{E_{\rm T}^{\rm miss}}}

\def\lsim{\mathrel{\raise.3ex\hbox{$<$\kern-.75em\lower1ex\hbox{$\sim$}}}}
\def\gsim{\mathrel{\raise.3ex\hbox{$>$\kern-.75em\lower1ex\hbox{$\sim$}}}}

\newcommand{\bea}{\begin{eqnarray}}
\newcommand{\eea}{\end{eqnarray}}

\newcommand{\be}{\begin{equation}}
\newcommand{\ee}{\end{equation}}

\title{Two Higgs Doublets and a Complex Singlet:\\
Disentangling the Decay Topologies and Associated Phenomenology
}

\author[a,b]{Sebastian~Baum,}
\author[c]{Nausheen~R.~Shah}

\affiliation[a]{The Oskar Klein Centre for Cosmoparticle Physics, Department of Physics, Stockholm University, Alba Nova, 10691 Stockholm, Sweden}
\affiliation[b]{Nordita, KTH Royal Institute of Technology and Stockholm University, Roslagstullsbacken 23, 10691 Stockholm, Sweden}
\affiliation[c]{Department of Physics \& Astronomy, Wayne State University, Detroit, MI 48201, USA}

\emailAdd{sbaum@fysik.su.se}
\emailAdd{nausheen.shah@wayne.edu}

\preprint{NORDITA-2018-067, 
WSU-HEP-1803}

\abstract{
We present a systematic study of an extension of the Standard Model (SM) with two Higgs doublets and one complex singlet (2HDM+S). In order to gain analytical understanding of the parameter space, we re-parameterize the 27 parameters in the Lagrangian by quantities more closely related to physical observables: physical masses, mixing angles, trilinear and quadratic couplings, and vacuum expectation values. Embedding the 125\,GeV SM-like Higgs boson observed at the LHC places stringent constraints on the parameter space. In particular, the mixing of the SM-like interaction state with the remaining states is severely constrained, requiring approximate alignment without decoupling in the region of parameter space where the additional Higgs bosons are light enough to be accessible at the LHC. In contrast to 2HDM models, large branching ratios of the heavy Higgs bosons into two lighter Higgs bosons or a light Higgs and a $Z$ boson, so-called Higgs cascade decays, are ubiquitous in the 2HDM+S. Using currently available limits, future projections, and our own collider simulations, we show that combining different final states arising from Higgs cascades would allow to probe most of the interesting region of parameter space with Higgs boson masses up to 1\,TeV at the LHC with $L=3000\,{\rm fb}^{-1}$ of data.
}

\begin{document}

\maketitle
\flushbottom

\section{Introduction}
With the discovery of a Higgs boson at the Large Hadron Collider (LHC) in 2012~\cite{Aad:2012tfa,Chatrchyan:2012xdj} with mass $m_{h_{\rm 125}} \approx 125\,$GeV~\cite{Aad:2015zhl,Sirunyan:2017exp} and couplings compatible with those of a Standard Model (SM) Higgs~\cite{Khachatryan:2016vau,CMS-PAS-HIG-16-042,CMS-PAS-HIG-17-031,Sirunyan:2017exp,ATLAS-CONF-2017-047}, the final ingredient of the SM has been found. However, despite the SM holding up to all laboratory test it has been subjected to so far, it fails to explain the behavior of the Universe at large scales. In particular, the SM does not contain a suitable candidate for the observed Dark Matter and fails to explain the matter-antimatter asymmetry. Beyond such phenomenological problems, the SM also suffers from some issues more theoretical in nature, e.g. the hierarchy problem or the lack of explanation of the SM's flavor structure.

During the past 50 years, a multitude of Beyond the Standard Model (BSM) particle physics models have been developed to address the aforementioned problems. The vast majority of BSM models, in particular those valid to energy scales much larger than the electroweak scale, feature a scalar sector extended beyond the SM's one $SU(2)$-doublet Higgs. For example, consistent supersymmetric extensions of the SM require a Higgs sector containing at least two Higgs doublets. Furthermore, while the parameters of the Higgs sector are those least well measured in the SM, many of the SM's shortcomings are intimately related to the scalar sector e.g. the hierarchy problem, the matter-antimatter asymmetry, and the flavor structure.

There are two avenues for studying BSM physics: One can either take a top down approach, starting from well-motivated SM extensions valid at energy scales much larger than the electroweak scale, or, one can take a bottom up approach by parameterizing our ignorance of high-scale physics by writing down the most general form of BSM models at the weak scale. In the case of extensions of the SM's Higgs sector, this may be exemplified by the well studied case of BSM models containing two Higgs doublets. One can either choose to study more complete models containing such a Higgs sector, e.g. the Minimal Supersymmetric Standard Model (MSSM), or, one can take a more model independent approach and study the most general form of a two Higgs doublet Model~(2HDM).

From a bottom up perspective, extensions of 2HDMs with an additional singlet may be motivated from the well known fact that they facilitate baryogenesis~\cite{Profumo:2007wc,Barger:2008jx,Cline:2012hg} to explain the matter-antimatter asymmetry, from flavor considerations~\cite{Campos:2017dgc,Camargo:2018klg} and from being useful for constructing Dark Matter models~\cite{Baum:2017enm}. In addition, well motivated top down BSM models exist containing such a scalar sector, for example the Next-to-Minimal Supersymmetric Standard Model (NMSSM)~\cite{Maniatis:2009re,Ellwanger:2009dp}. In this work, we take the bottom up approach to study extensions of the SM's scalar sector with one additional doublet and one complex SM gauge singlet $S$, which we refer to as 2HDM+S in the following.

While a large body of literature exists on general 2HDM models, see e.g. Refs.~\cite{Branco:2011iw,Bernon:2015qea,Bernon:2015wef,Basler:2017nzu,Cherchiglia:2017uwv} for recent works, no systematic study of 2HDM+S models exists in the literature to the best of our knowledge. Some aspects of 2HDM+S models have been discussed in Ref.~\cite{Chalons:2012qe} and in the appendix of Ref.~\cite{Carena:2015moc}. The extension of 2HDMs with a real singlet has been discussed in Refs.~\cite{Chen:2013jvg, vonBuddenbrock:2016rmr, Muhlleitner:2016mzt, Muhlleitner:2017dkd, vonBuddenbrock:2017gvy}. In this work, we provide a first systematic study of the 2HDM+S parameter space. The physical Higgs sector contains 5 real neutral Higgs bosons and one charged Higgs boson after electroweak symmetry breaking. Assuming CP-conservation, the 5 neutral Higgs bosons can be subdivided into 3 CP-even states and 2 CP-odd states. In order to be compatible with the observed phenomenology, one of the CP-even states must have couplings to pairs of SM particles similar to those of a SM Higgs with a mass of $125\,$GeV, such that it can be identified with the 125\,GeV Higgs boson observed at the LHC. As we will see, the presence of this SM-like state has important implications for the behavior of the remaining Higgs bosons. 

Extended Higgs sectors such as the 2HDM+S can be tested at the LHC in two complimentary ways: one can study the behavior of the observed approximately SM-like $h_{125}$, constraining its possible mixing with additional Higgs bosons, or, one may directly search for the remaining Higgs bosons in the model. The first way has thus far constrained a number of couplings of $h_{125}$ to SM particles to be within $\mathcal{O}(10\,\%)$ of the SM values. In most cases, future data expected to be collected at the LHC during the next 20 years will allow only for marginal improvements on these bounds~\cite{Englert:2015hrx}. Note however, that a Higgs factory such as the International Linear Collider would allow for precision tests of the SM-like Higgs boson, probing some of its couplings below the $\mathcal{O}(1\,\%)$ level~\cite{Asner:2013psa}. This would yield powerful constraints on any BSM Higgs sector. 

Such indirect searches are complimented by direct searches for the additional Higgs bosons as carried out at the LHC in a plenitude of final states. Most commonly, one searches for direct production of the additional Higgs bosons which then decay into pairs of SM particles. Such search strategies are similar to those employed in the hunt for the 125\,GeV Higgs boson prior to its discovery. For a 2HDM Higgs sector, searches for additional Higgs bosons are complicated by three different issues: 1) While CP-even states are allowed to decay into pairs of SM vector bosons, the SM-like nature of $h_{125}$ suppresses the couplings of the additional states to pairs of vector bosons.~\footnote{Here and in the rest of the article, vector bosons refers specifically to the weak gauge bosons: $W$ and $Z$.} 2) In addition to suppressing the branching ratio into pairs of SM vector bosons, the SM-like nature of $h_{125}$ also suppresses the coupling of heavy Higgs bosons to pairs of SM-like Higgs boson or a $Z$ boson and a SM-like Higgs, reducing the power of the corresponding search channels. 3) For large regions of parameter space the decays of any Higgs boson with mass $m_\Phi \gtrsim 350\,$GeV are dominated by decays into pairs of top quarks. Due to the interference of this signal with the QCD background~\cite{Dicus:1994bm,Barcelo:2010bm,Barger:2011pu,Bai:2014fkl,Jung:2015gta,Craig:2015jba,Gori:2016zto,Carena:2016npr}, such decays are challenging to use for Higgs searches at the LHC with current search strategies. 

All three of these complications are still present in models with enlarged Higgs sectors such as the 2HDM+S. In addition, the presence of the singlet, which can mix with the doublets, complicates conventional searches further since the production cross section of any Higgs state at the LHC is suppressed with a growing singlet fraction. However, the presence of additional Higgs bosons allows for new decay channels of the heavy Higgs bosons, so-called {\it Higgs cascades}, where the heavy Higgs bosons decay into two lighter Higgs states or a light Higgs and a $Z$ boson~\cite{Kang:2013rj,King:2014xwa,Carena:2015moc,Ellwanger:2015uaz,Costa:2015llh,Baum:2017gbj,Ellwanger:2017skc}. Such decays can also be present in the generic 2HDM without the singlet if the CP-even and CP-odd Higgs bosons have non-degenerate masses, although large mass splittings are difficult to achieve in consistent 2HDMs~\cite{Krauss:2018thf}. However, in this case the direct production cross section of the lighter state has no suppression from a singlet component, generally making direct searches for such a state more powerful. We briefly discuss the limiting case of decoupling the singlet from the doublets in Sec.~\ref{sec:xsec_BR}. 

As mentioned above, the branching ratios of any of the heavy Higgs bosons to pairs of SM-like $h_{125}$'s, pairs of vector bosons, or a $Z$ boson and a $h_{125}$, are suppressed by the SM-like nature of $h_{125}$. Of the remaining decay modes with possibly large, $\mathcal{O}(10\,\%)$, branching ratios, those consisting of a SM-like Higgs boson or a $Z$ boson and one additional non SM-like Higgs are most useful for searches at the LHC. This is because one can use the decay products of the SM-like Higgs or the $Z$ boson with known mass and branching ratios to tag such events. Furthermore, final state search signatures will also contain the decay products of the non SM-like Higgs bosons: if they decay into visible states, e.g. a pair of SM particles, one searches for a resonance in the invariant mass spectrum of the decay products. However, the non SM-like Higgs bosons can also decay into additional new states, for example pairs of new stable neutral particle playing the role of a Dark Matter candidate. In this case, the decay products of the non SM-like Higgs bosons produced in the Higgs cascades would leave the detector without depositing energy, manifesting as {\it missing transverse energy} (\MET) in the detector, giving rise to so-called {\it mono-$Z$} and {\it mono-Higgs} signatures. 

As shown in the context of the NMSSM in Refs.~\cite{Carena:2015moc,Baum:2017gbj,Ellwanger:2017skc}, the branching ratios of Higgs cascades are sizable in large parts of the parameter space, leading to observable signatures at the LHC. As we will see below, this is also the case in the general 2HDM+S.

The remainder of this work is organized as follows: In Sec.~\ref{sec:2HDMS}, we discuss the 2HDM+S. We focus on the mass spectrum, and the implications of the SM-like nature of the 125\,GeV Higgs boson in Sec.~\ref{sec:2HDMS_mass}, and the couplings between the Higgs bosons and the relevant model parameters for the Higgs cascades in Sec.~\ref{sec:2HDMS_tricoup}. In Sec.~\ref{sec:xsec_BR} we discuss the production cross sections and decay modes of the 2HDM+S Higgs bosons, and in Sec.~\ref{sec:cascaderatios} we discuss the Higgs cascades in more detail. The more experimentally inclined reader may skip directly to Sec.~\ref{sec:LHC} where we discuss the future reach of the LHC for the full 2HDM+S using Higgs cascades. 
In Sec.~\ref{sec:LHC_2HDMS} we demonstrate that the 2HDM+S gives rise to cross sections within reach of the Higgs cascade searches at the LHC for large regions of parameter space. We reserve Sec.~\ref{sec:conclusions} for our conclusions. Appendices~\ref{app:2HDMS_mass}, \ref{app:2HDMS_tricoup}, \ref{app:2HDMS_quartics}, \ref{app:monoZsim}, and \ref{app:NMSSMmap} contain details about the mass matrices, trilinear couplings, quartic couplings, the collider simulation performed for estimating the reach of the mono-$Z$ channel, and the mapping of the 2HDM+S parameters onto the NMSSM, respectively. 

\section{2HDM+S}\label{sec:2HDMS}
The most general 2HDM scalar potential is given by~\cite{Branco:2011iw}
\begin{equation} \begin{split} \label{eq:V2HDM}
	V_{\rm 2HDM} &= m_{11}^2 \Phi_1^\dagger \Phi_1 + m_{22}^2 \Phi_2^\dagger \Phi_2 - \left( m_{12}^2 \Phi_1^\dagger \Phi_2 + {\rm h.c.} \right) \\
	& \qquad+ \frac{\lambda_1}{2} \left( \Phi_1^\dagger \Phi_1 \right)^2 + \frac{\lambda_2}{2} \left( \Phi_2^\dagger \Phi_2 \right)^2 + \lambda_3 \left( \Phi_1^\dagger \Phi_1 \right) \left( \Phi_2^\dagger \Phi_2 \right) + \lambda_4 \left( \Phi_1^\dagger \Phi_2 \right) \left( \Phi_2^\dagger \Phi_1 \right) \\
	& \qquad+ \left[ \frac{\lambda_5}{2} \left( \Phi_1^\dagger \Phi_2 \right)^2 + \lambda_6 \left( \Phi_1^\dagger \Phi_1 \right) \left( \Phi_1^\dagger \Phi_2 \right) + \lambda_7 \left( \Phi_2^\dagger \Phi_2 \right) \left( \Phi_1^\dagger \Phi_2 \right) + \rm{h.c.} \right],
\end{split} \end{equation}
where $\Phi_1$, $\Phi_2$ are $SU(2)$ doublets with hypercharge $Y=1/2$. The $m_{ij}^2$ parameters have dimension mass squared, while the $\lambda_i$ are dimensionless. In this work, we study the CP-conserving case, where all parameters can be chosen manifestly real. Extending the field content by a complex scalar SM gauge singlet $S$, the most general scalar potential for the additional terms is
\begin{equation} \begin{split} \label{eq:VS}
	V_S &= \left( \xi S + {\rm h.c.} \right) + m_S^2 S^\dagger S + \left( \frac{m_S'^2}{2} S^2 + {\rm h.c.} \right) \\
	& \qquad+ \left( \frac{\mu_{S1}}{6} S^3 + {\rm h.c.} \right) + \left( \frac{\mu_{S2}}{2} S S^\dagger S + {\rm h.c.} \right) \\
	& \qquad+ \left( \frac{\lambda''_1}{24} S^4 + {\rm h.c.} \right) + \left( \frac{\lambda''_2}{6} S^2 S^\dagger S + {\rm h.c.} \right) + \frac{\lambda''_3}{4} \left( S^\dagger S \right)^2 \\
	& \qquad+ \left[ S \left( \mu_{11} \Phi_1^\dagger \Phi_1 + \mu_{22} \Phi_2^\dagger \Phi_2 + \mu_{12} \Phi_1^\dagger \Phi_2 + \mu_{21} \Phi_2^\dagger \Phi_1 \right) + {\rm h.c.} \right] \\
	& \qquad+ S^\dagger S \left[ \lambda'_1 \Phi_1^\dagger \Phi_1 + \lambda'_2 \Phi_2^\dagger \Phi_2 + \left( \lambda'_3 \Phi_1^\dagger \Phi_2 + {\rm h.c.} \right) \right] \\ 
	& \qquad+ \left[ S^2 \left( \lambda'_4 \Phi_1^\dagger \Phi_1 + \lambda'_5 \Phi_2^\dagger \Phi_2 + \lambda'_6 \Phi_1^\dagger \Phi_2 + \lambda'_7 \Phi_2^\dagger \Phi_1 \right) + {\rm h.c.} \right].
\end{split} \end{equation}
The parameter $\xi$ has dimensions mass cubed, the parameters $\{\mu_{Si}, \mu_{ij}\}$ have dimension mass, and the $\{\lambda'_i, \lambda''_i\}$ are dimensionless. For the CP-conserving case, all parameters can again be chosen to be manifestly real. 

The most general scalar potential for a 2HDM+S model is then given by
\begin{equation} \label{eq:V}
	V = V_{2{\rm HDM}} + V_S \;.
\end{equation}
Note that the NMSSM has the same Higgs sector, although supersymmetry severely restricts the parameters. We provide a mapping of the 2HDM+S onto the Higgs sector of the NMSSM in Appendix~\ref{app:NMSSMmap}.

Out of the 29 parameters in the scalar potential only 27 are physical. One can perform a global $SO(2)$ rotation in $(\Phi_1, \Phi_2)$ space~\cite{Branco:2011iw}, and a real shift of $S$. These transformations render one of the parameters in the doublet sector and one parameter in the singlet sector unphysical, respectively. In the following, we will use the basis where the shift of $S$ has been used to eliminate the tadpole term $\xi$ from the scalar potential. We do not further define the particular basis for the doublets yet, as such a choice is intimately linked to the choice of couplings to the SM fermions, discussed below.

One can furthermore use the minimization conditions 
\begin{equation}
	\left.\frac{\partial V}{\partial \Phi_1}\right|_{\substack{\Phi_1 = v_1\\\Phi_2 = v_2\\S=v_S}} = \left.\frac{\partial V}{\partial \Phi_2}\right|_{\substack{\Phi_1 = v_1\\\Phi_2 = v_2\\S=v_S}} = \left.\frac{\partial V}{\partial S}\right|_{\substack{\Phi_1 = v_1\\\Phi_2 = v_2\\S=v_S}} = 0 \;,
\end{equation}
to trade three parameters, e.g. $\{m_{11}^2, m_{22}^2, m_S^2\}$, for the vacuum expectation values (vevs)
\begin{equation}
	v_1 \equiv \left\langle \Phi_1\right\rangle, \qquad v_2 \equiv \left\langle \Phi_2\right\rangle, \qquad v_S \equiv \left\langle S\right\rangle.
\end{equation}
For the purposes of this work, we will assume that the vevs lie along the neutral direction of the doublets and that the vacuum is CP-conserving. Note that these assumptions may be broken even when using the vevs $\{v_1,v_2,v_S\}$ as input parameters and choosing them manifestly real, as we will do in the following, see for example the discussion in Ref.~\cite{Branco:2011iw}. Further, deeper minima, including CP-violating ones, may be present. An analysis of the vacuum structure of the 2HDM+S is beyond the scope of this work and we leave it for the future.

In addition to the field transformations discussed above, one can redefine each of the doublets as well as the singlet by a phase. Demanding the potential to remain manifestly CP-conserving under such transformations, i.e. all parameters  remain real, constrains these transformation to $U(1) \otimes Z_2$ in the doublet sector and $Z_2$ for the singlet. Although these transformations cannot be employed to absorb additional parameters, one can use them to enforce all vevs positive, $\{v_1, v_2, v_S\} \geq 0$. 

As customary, we define
\begin{equation}
	v \equiv \sqrt{v_1^2 + v_2^2} \;, \qquad \tan\beta \equiv v_1/v_2 \;.
\end{equation}
The observed mass of the $Z$ boson $m_Z = 91.2\,$GeV is obtained for $v = 174\,$GeV.

It is useful to rotate the Higgs fields to the {\it extended Higgs basis}~\cite{Georgi:1978ri, Donoghue:1978cj, gunion2008higgs, Lavoura:1994fv, Botella:1994cs, Branco99, Gunion:2002zf,Carena:2015moc}\footnote{Note that there are different conventions in the literature for the Higgs basis differing by an overall sign of $H^{\rm NSM}$ and $A^{\rm NSM}$. Taking these into account, our potential and couplings for the 2HDM+S can be mapped directly to the potential and couplings given in the appendices of Ref.~\cite{Carena:2015moc}.}
\begin{align} \label{eq:Hbasis1}
	\begin{bmatrix} G^+ \\ \frac{1}{\sqrt{2}} \left(H^{\rm SM} + i G^0\right) \end{bmatrix} &= \sin\beta \Phi_1 + \cos\beta \Phi_2 \;, \\
	\begin{bmatrix} H^+ \\ \frac{1}{\sqrt{2}} \left(H^{\rm NSM} + i A^{\rm NSM}\right) \end{bmatrix} &= \cos\beta \Phi_1 - \sin\beta \Phi_2 \;, \\
	\frac{1}{\sqrt{2}} \left( H^{\rm S} + i A^{\rm S} \right) &= S \;,
	\label{eq:Hbasis-1}
\end{align} 
where $\{H^{\rm SM}, H^{\rm NSM}, H^{\rm S}\}$ and $\{A^{\rm NSM}, A^{\rm S}\}$ are the neutral CP-even and CP-odd real Higgs basis interaction states and $G^0$ ($G^\pm$) is the neutral (charged) Goldstone mode. In this basis, of the states coming from the doublets, only $\left\langle H^{\rm SM}\right\rangle = \sqrt{2} v$ acquires a vev, and it is straightforward to work out the coupling of SM fermions to the Higgs basis states to e.g. study possible flavor changing neutral currents (FCNCs). Potentially dangerous FCNCs can be omitted if the groups of right-handed SM fermions with the same quantum numbers couple to only one of the doublets, respectively, as in the so-called {\it Type I}, {\it Type II}, {\it flipped}, and {\it lepton specific} 2HDM versions~\footnote{In Type I models, all SM fermions couple to the same doublet. In Type II models, one doublet couples to up-type fermions and the other doublet to the down-type fermions. In flipped models, the up-type quarks and the charged leptons couple to one of the doublets while down-type quarks couple to the other doublet. Finally, in lepton specific models, all quarks couple to one doublet while the charged leptons couple to the other doublet.}~\cite{Branco:2011iw}. Note that couplings of the fermions to the singlet $S$ are forbidden by gauge invariance. Restricting ourselves to the case where the values of the Yukawa matrices are chosen such that the observed SM fermion masses are obtained, the couplings of pairs of SM particles to the Higgs basis states can be written as 
\begin{align}
	H^{\rm SM} \left( f_1, f_2, {\rm VV} \right) &= \left( g_{\rm SM}, \; g_{\rm SM}, \; g_{\rm SM} \right), 
	\\ H^{\rm NSM} \left( f_1, f_2, {\rm VV} \right) &= \left( g_{\rm SM} / \tan\beta, \; - g_{\rm SM} \tan\beta, \; 0 \right),
	\\ H^{\rm S} \left( f_1, f_2, {\rm VV} \right) &= \left( 0, \; 0, \; 0 \right),
	\\ A^{\rm NSM} \left( f_1, f_2, {\rm VV} \right) &= \left( g_{\rm SM} / \tan\beta, \; g_{\rm SM} \tan\beta, \; 0 \right), 
	\\ A^{\rm S} \left( f_1, f_2, {\rm VV} \right) &= \left( 0, \; 0, \; 0 \right), 
\end{align}
where ``$f_1$'' (``$f_2$'') stands for SM-fermions coupling to $\Phi_1$ ($\Phi_2$), ``VV'' for pairs of $W$ or $Z$ gauge bosons, and $g_{\rm SM}$ is the coupling of a SM Higgs boson to such particles. Note that CP-odd states couple to pseudoscalar fermion bilinears $(\bar{f} \gamma_5 f)$, instead the CP-even states couple to the scalar bilinear $(\bar{f} f)$.

For concreteness, in the following we assume a Type~II Yukawa structure,
\begin{equation} \label{eq:Yuk}
	\mathcal{L}_{\rm Yuk} = - Y_u \bar{Q} \cdot \Phi_1 u_R - Y_d \bar{Q} \cdot \widetilde{\Phi}_2 d_R - Y_e \bar{L} \cdot \widetilde{\Phi}_2 e_R \;, 
\end{equation} 
where $\widetilde{\Phi} \equiv i \sigma_2 \Phi^*$, the $Y_i$ are the $3 \times 3$ Yukawa matrices, and the left-handed quarks $Q$ and leptons $L$ as well as the right-handed up-type (down-type) quarks $u_R$ ($d_R$) and the right-handed leptons $e_R$ should be understood as vectors in generation space. We define all parameters of the 2HDM+S scalar potential, in particular the vevs, in the interaction basis as determined by the Yukawa structure~[cf. the discussion below Eq.~\eqref{eq:V}], and where the singlet is shifted such that $\xi = 0$. In particular, this means that we keep all parameters of the doublet sector in our expressions, but one should keep in mind that one of these parameters is unphysical due to the $SO(2)$ symmetry in $(\Phi_1, \Phi_2)$ space.

We stress that the scalar potential of our model allows more general Yukawa structures than the one chosen in Eq.~\eqref{eq:Yuk}, including ones leading to tree-level FCNCs. Consistency with a desired Yukawa structure, e.g.  Type I or Type II, can be ensured by imposing an ad-hoc $Z_2$ symmetry, as is done in generic 2HDM's. For example, the Type II structure in Eq.~\eqref{eq:Yuk} and its radiative stability can be ensured by a $Z_2$ symmetry under which $\Phi_1 \to - \Phi_1$, $u_R \to - u_R$, and all other fields transform trivially.\footnote{Ensuring a Type II Yukawa structure via this particular implementation of a $Z_2$ symmetry would enforce $\lambda_6 = \lambda_7 = \lambda_3' = \lambda_6' = \lambda_7' = 0$ in the scalar potential. Note, that the dimensionful parameters $m_{12}^2$, $\mu_{12}$, and $\mu_{21}$ may still be different from zero corresponding to a soft breaking of the $Z_2$ symmetry.} In this work, we remain agnostic about the mechanism which ensures the desired Yukawa structure and its radiative stability. Hence, we will not impose additional symmetries on the model. The inclined reader may choose their favorite mechanism; imposing the corresponding restrictions on the 2HDM+S parameter space is straightforward. Note also that our results in the remainder of this paper will in general hold  for a different Yukawa structure. Some quantitative details may change because of the change of the Yukawa enhancement/suppression of the fermion couplings. However, such modifications will be small since we mostly consider the low $\tan\beta = \mathcal{O}(1)$ regime in this work. 

\subsection{Higgs Mass Eigenstates and Alignment} \label{sec:2HDMS_mass}
The mass eigenstates are obtained from the diagonalization of the squared-mass matrix for the Higgs basis states. The neutral and charged Goldstone modes $G^0$ and $G^\pm$ are massless by construction and do not mix with the other interaction states. In the following, we remove the Goldstone modes from the theory by choosing the unitary gauge. Furthermore, there is no mixing between CP-even and CP-odd states in the CP-conserving 2HDM+S. 

We denote the three CP-even mass eigenstates 
\begin{equation}
h_i = \{ h_{125}, H, h\}\;, 
\end{equation}
where $h_{125}$ is identified with the $m_{h_{125}} \approx 125\,$GeV SM-like state observed at the LHC, and $H$ and $h$ are ordered by masses, $m_H > m_h$. Each mass eigenstate is an admixture of the extended Higgs basis interaction states,
\begin{equation}
	h_i = S_{h_i}^{\rm SM} H^{\rm SM} + S_{h_i}^{\rm NSM} H^{\rm NSM} + S_{h_i}^{\rm S} H^{\rm S} \;,
\end{equation}
where $S_{h_i}^{j}$ with $j = \{$SM, NSM, S$\}$ denotes the components of the mass eigenstates in terms of the interaction basis. Likewise, we denote the two CP-odd mass eigenstates 
\begin{equation}
a_i = \{A, a\}\;,
\end{equation}
where again $m_A > m_a$, and
\begin{equation}
	a_i = P_{a_i}^{\rm NSM} A^{\rm NSM} + P_{a_i}^{\rm S} A^{\rm S} \;,
\end{equation}
where the components are similarly denoted by $P_{a_i}^{j}$.

The $S_{h_i}^j$ ($P_{a_i}^j$) are obtained from diagonalizing the (symmetric) squared mass matrix for the CP-even (CP-odd) Higgs bosons, $\mathcal{M}_{S}^2$ ($\mathcal{M}_{P}^2$). The values of the entries of the mass matrices and the mass of the charged Higgs boson are recorded in Appendix~\ref{app:2HDMS_mass}.

The observation of a $m_{h_{125}} \approx 125\,$GeV mass eigenstate with couplings to SM particles compatible with that of a SM Higgs boson at the LHC implies that our model must contain a mass eigenstate with
\begin{equation}
	m_{h_{125}} \approx 125\,{\rm GeV}\;, \qquad S_{h_{125}}^{\rm SM} \approx 1 \;, \qquad \{ (S_{h_{125}}^{\rm NSM})^2, (S_{h_{125}}^{\rm S})^2 \} \ll 1\;,
\end{equation} 
or, in other words, a 125\,GeV mass eigenstate approximately {\it aligned} with the $H^{\rm SM}$ interaction state.

The observed mass approximately fixes $\mathcal{M}_{S,11}^2 \approx m_{h_{125}}^2 \approx \left(125\,{\rm GeV}\right)^2$, while the mixing of $H^{\rm SM}$ with $H^{\rm NSM}$ is suppressed~[$(S_{h_{125}}^{\rm NSM})^2 \ll 1$] for 
\begin{equation} \label{eq:DDmix}
	| \mathcal{M}_{S,12}^2 | \ll | \mathcal{M}_{S,22}^2 - \mathcal{M}_{S,11}^2 | \;,
\end{equation}
and the mixing of $H^{\rm SM}$ with $H^{\rm S}$ is suppressed~[$(S_{h_{125}}^{\rm S})^2 \ll 1$] for
\begin{equation} \label{eq:DSmix}
	| \mathcal{M}_{S,13}^2 | \ll | \mathcal{M}_{S,33}^2 - \mathcal{M}_{S,11}^2 | \;.
\end{equation}
Hence we see that there are two possibilities to achieve alignment: either the left hand sides of Eqs.~\eqref{eq:DDmix} and ~\eqref{eq:DSmix} go to zero, or, the right hand sides become large while the left hand sides remain non-zero and sizable. The latter possibility is the so-called {\it decoupling} limit, corresponding to
\begin{equation}
	\left\{ |\mathcal{M}_{S,22}^2|, |\mathcal{M}_{S,33}^2| \right\} \gg \mathcal{M}_{S,11}^2 \approx \left( 125\,{\rm GeV} \right)^2 ~,
\end{equation}
implying $\{m_H, m_h \} \gg m_{h_{125}}$. The first option is the so-called {\it alignment without decoupling} limit, and is of particular interest for LHC phenomenology. This is because the additional CP-even mass eigenstates $H$ and $h$ are not necessarily much heavier than $h_{125}$ and hence may be directly accessible at the LHC~\cite{Gunion:2002zf, Carena:2013ooa, Dev:2014yca, Carena:2014nza, Bernon:2015qea, Bernon:2015wef, Carena:2015moc}. In this case,
\begin{equation}
	|\mathcal{M}_{S,1i}^2| \ll | \mathcal{M}_{S,ii}^2-\mathcal{M}_{S,11}^2 | \ll \left( 125\,{\rm GeV} \right)^2\; ; \qquad i=\{2,3\},
\end{equation}
must be satisfied in order to ensure approximate alignment. Neglecting radiative corrections, perfect alignment is achieved for~\footnote{Recall that we use the interaction basis as defined by the Yukawa structure, Eq.~\eqref{eq:Yuk}, and where the singlet field $S$ is shifted such that the tadpole term $\xi = 0$. Please see the Appendices of Ref.~\cite{Carena:2015moc} for the alignment conditions given in terms of parameters defined in the extended Higgs basis of the 2HDM+S.} 
\begin{align} \label{eq:align1}
	\lambda_3 + \lambda_4 + \lambda_5 &= - \frac{1}{c_{2\beta}} \left( \lambda_1 s_\beta^2 - \lambda_2 c_\beta^2 + \frac{\lambda_6 s_{3\beta}}{c_\beta} + \frac{\lambda_7 c_{3\beta}}{s_\beta} \right) , 
	 \\ 
	 -\left(\mu_{12} + \mu_{21}\right) &= \frac{\mu_{11} t_\beta^2 + \mu_{22}}{t_\beta} + v_S \left[ \left( \lambda'_1 + 2 \lambda'_4 \right) t_\beta + \frac{\lambda'_2 + 2 \lambda'_5}{t_\beta} + 2\left( \lambda'_3 + \lambda'_6 + \lambda'_7 \right) \right] , \label{eq:align2}
\end{align} 
where the first condition ensures $\mathcal{M}_{S,12}^2 = 0$ and the second condition $\mathcal{M}_{S,13}^2 = 0$. Here and in the following we employ a shorthand notation, 
\begin{equation}
	s_\beta \equiv \sin\beta\;, \qquad c_\beta \equiv \cos\beta \;, \qquad t_\beta \equiv \tan\beta \;.
\end{equation} 

\subsection{Couplings and Parameters} \label{sec:2HDMS_tricoup}
The trilinear couplings of the Higgs bosons can be obtained from the scalar potential as
\begin{equation}
	g_{\Phi_i \Phi_j \Phi_k} \equiv -\left.\frac{\partial^3 \mathcal{L}}{\partial \Phi_i \partial \Phi_j \partial \Phi_k}\right|_{\substack{\Phi_1 = v_1\\\Phi_2 = v_2\\S = v_S}} = \left.\frac{\partial^3 V}{\partial \Phi_i \partial \Phi_j \partial \Phi_k}\right|_{\substack{\Phi_1 = v_1\\\Phi_2 = v_2\\S = v_S}} \;.
\end{equation}
We first note that CP-conservation forbids couplings such as $g_{h_i h_j a_k}$ or $g_{a_i a_j a_k}$.
In the following, we discuss the remaining trilinear couplings in the extended Higgs basis. From these, the couplings for the mass eigenstates can be obtained via
\begin{equation} \begin{split} \label{eq:tricoup_mass}
	g_{h_i h_j h_k} &= \sum_{H^l} \sum_{H^m} \sum_{H^n} S_{h_i}^{H^l} S_{h_j}^{H^m} S_{h_k}^{H^n} g_{H^l H^m H^n} \;, \\
	g_{h_i a_j a_k} &= \sum_{H^l} \sum_{A^m} \sum_{A^n} S_{h_i}^{H^l} P_{a_j}^{A^m} P_{a_k}^{A^n} g_{H^l A^m A^n}\;.
\end{split} \end{equation}

Some of the couplings allowed by CP-conservation, vanish due to gauge invariance. For example, the coupling 
\begin{equation}
	g_{H^{\rm NSM} A^{\rm NSM} A^{\rm S}} = 0\;,
\end{equation}
is identically zero since it would arise only through terms proportional to $S(\Phi_i \Phi_j)$, $i,j=\{1,2\}$, in the scalar potential which are forbidden by $U(1)_Y$ invariance.

A number of the trilinear couplings are proportional to entries in the mass matrices: Categorizing the Higgs basis states as SM-like ($H^{\rm SM}$), NSM-like ($H^{\rm NSM}, A^{\rm NSM}, H^\pm$), and singlet-like ($H^{\rm S}, A^{\rm S}$), we find that couplings involving only SM-like states are proportional to the corresponding diagonal entry of the CP-even mass matrix
\begin{equation}
	({\rm SM-SM-SM}) \propto \mathcal{M}_{S,11}^2 \;.
\end{equation}
Couplings involving two SM-like states are proportional to the corresponding entry in the CP-even squared mass matrix mixing such states,
\begin{align}
	({\rm SM-SM-NSM}) \propto \mathcal{M}_{S,12}^2 \;,
	\\ ({\rm SM-SM-S}) \propto \mathcal{M}_{S,13}^2 \;.
\end{align}
Couplings involving one SM-like state, one NSM-like state, and one singlet-like state are proportional to the singlet-doublet mixing entries in one of the mass matrices
\begin{align}
	({\rm SM-NSM}-H^{\rm S}) \propto \mathcal{M}_{S,23}^2 \;,
	\\ ({\rm SM-NSM}-A^{\rm S}) \propto \mathcal{M}_{P,12}^2 \;.
\end{align}

This implies that the couplings of two SM-like states to one NSM-like or one singlet-like state are suppressed in the alignment limit. Indeed, for perfect alignment one finds
\begin{equation}
	\{ g_{H^{\rm SM} H^{\rm SM} H^{\rm NSM}}, \quad g_{H^{\rm SM} H^{\rm SM} H^{\rm S}} \} \to 0 \;,
\end{equation}
and hence the couplings of the mass eigenstates,
\begin{equation}
	g_{h_{125}h_{125}H}\;, \quad g_{h_{125}h_{125}h} \;,
\end{equation}
are suppressed in proximity to alignment, and vanish for perfect alignment. 

Besides the trilinear couplings proportional to the entries of the mass matrices, a few of the couplings are identical or can be written as linear combinations of entries of the mass matrices. Accounting for such degeneracies, we find that there are 10 independent trilinear couplings, 
\begin{equation}\label{trilin} \begin{split}
	&\{ g_{H^{\rm SM} H^{\rm NSM} H^{\rm NSM}}, \quad g_{H^{\rm SM} H^{\rm S} H^{\rm S}}, \quad g_{H^{\rm SM} A^{\rm S} A^{\rm S}} \}, \\
	&\{ g_{H^{\rm NSM} H^{\rm NSM} H^{\rm NSM}}, \quad g_{H^{\rm NSM} H^{\rm NSM} H^{\rm S}}, \quad g_{H^{\rm NSM} H^{\rm S} H^{\rm S}}, \quad g_{H^{\rm NSM} A^{\rm S} A^{\rm S}} \}, \\
	&\{ g_{H^{\rm S} H^{\rm S} H^{\rm S}}, \quad g_{H^{\rm S} A^{\rm NSM} A^{\rm S}}, \quad g_{H^{\rm S} A^{\rm S} A^{\rm S}} \} \;.
\end{split} \end{equation}
For completeness we provide a list of the trilinear couplings and their values not a priori forbidden by CP or charge conservation in Table~\ref{tab:2HDMStricoup} located in Appendix~\ref{app:2HDMS_tricoup}, which match the results obtained in Ref.~\cite{Carena:2015moc}.

It is straightforward to show from the scalar potential that there are 4 independent quartic couplings between the interaction states which cannot be written as linear combinations of the entries of the mass matrices or the trilinear couplings,
\begin{equation}\label{quartic}
	\{ \lambda_{H^{\rm NSM} H^{\rm NSM} H^{\rm S} H^{\rm S}}, \quad \lambda_{H^{\rm NSM} H^{\rm NSM} A^{\rm S} A^{\rm S}}, \quad \lambda_{H^{\rm S} H^{\rm S} A^{\rm S} A^{\rm S}} , \quad \lambda_{A^{\rm S} A^{\rm S} A^{\rm S} A^{\rm S}} \} .
\end{equation}
We list the values of these couplings in Appendix~\ref{app:2HDMS_quartics}.

All together, the 9 entries of the mass matrices, which we parameterize by the 5 physical masses
\begin{equation}
	m_{h_{125}}\;, \quad m_H\;, \quad m_{h}\;, \quad m_A\;, \quad m_{a}\;,
\end{equation} 
and 4 mixing angles
\begin{equation}
	\quad S_{h_{125}}^{\rm NSM}\;, \quad S_{h_{125}}^{\rm S}\;, \quad S_H^{\rm S}\;, \quad P_A^{\rm S}\;; 
\end{equation}
the 3 vevs, which we parameterize by 
\begin{equation}
	v\;, \quad \tan\beta\;, \quad v_S\;;
\end{equation}
the charged Higgs mass
\begin{equation}
	m_{H^\pm}\;;
\end{equation}
the 10 independent linear couplings listed in Eq.~\ref{trilin}, and the 4 independent quartic couplings given in Eq.~\ref{quartic} yield a set of 27 independent parameters. In the remainder of this work we will describe the 2HDM+S parameter space in terms of these parameters since they are more closely related to physical observables instead of the 27 parameters in the scalar potential. That said, we note that of these 27 parameters, only a subset are relevant for the analysis of the heavy Higgs decay topologies, especially given the constraints due to the SM-like nature of $h_{125}$. In particular, the quartics play no role~(however, these may be interesting when considering di-Higgs production), and as we shall see, only a few of the trilinear couplings listed in Eq.~\eqref{trilin} will be  needed. 

\subsection{Production Cross Section and Partial Decay Widths} \label{sec:xsec_BR}
For moderate/low values of $\tan\beta = \mathcal{O}(1)$, the production cross section for non-standard Higgs bosons at the LHC is dominated by gluon fusion. For larger values of $\tan\beta$, $bbH/bbA$ associated production may become relevant. Note that in proximity to the alignment limit the vector boson fusion production cross sections of additional Higgs boson is suppressed since only $H^{\rm SM}$ couples to pairs of vector bosons.

To first approximation, the gluon fusion production cross section can be parameterized by the contribution from top quarks in the loop. At one-loop order, the ratio of the gluon fusion production cross sections of a CP-even Higgs boson $h_i$ and a CP-odd Higgs boson $a_i$ is then given by~\cite{Spira:1995rr}
\begin{equation} \label{eq:ggf}
	\frac{\sigma(gg\to h_i)}{\sigma(gg\to a_j)} = \frac{\sigma_{ggh}(m_{h_i})}{\sigma_{ggh}(m_{a_i})} \left(\frac{S_{h_i}^{\rm SM} t_\beta - S_{h_i}^{\rm NSM}}{P_{a_i}^{\rm NSM}}\right)^2 \left[ \frac{1}{f(\tau_{a_i})} + \frac{\tau_{a_i}-1}{\tau_{a_i}} \right]^2 \;,
\end{equation}
where $\sigma_{ggh}(m)$ is the gluon fusion production cross section of a SM Higgs boson with mass $m$, $\tau_{a_i} \equiv (m_{a_i} / 2 m_t)^2$, $m_{h_i}$ is the mass of $h_i$, $m_{a_i}$ is the mass of $a_i$, $m_t$ is the top quark mass, and
\begin{equation}
	f(\tau_{a_i}) = \begin{cases} \arcsin^2\sqrt{\tau_{a_i}} &, \quad \tau_{a_i} \leq 1 \;,
	\\ -\frac{1}{4} \left[ \log\left(\frac{1+\sqrt{1-1/\tau_{a_i}}}{1-\sqrt{1-1/\tau_{a_i}}}\right) - i\pi \right]^2 &, \quad \tau_{a_i} >1 \;. \end{cases}
\end{equation}

As long as the narrow width approximation is valid, the production cross section of any final state arising from gluon fusion production of a Higgs boson is given by the product of the gluon fusion production cross section of the Higgs boson of interest, and its branching ratio into the relevant final state. The relevant branching ratio is given in terms of the partial decay widths by
\begin{equation}
	{\rm BR}(\Phi_i \to {\rm final\;state}) = \frac{\Gamma(\Phi_i \to {\rm final\;state})}{\Gamma_{\Phi_i}}\;,
\end{equation} 
where the width $\Gamma_{\Phi_i}$ of the state $\Phi_i$ is obtained by summing over all partial widths. In the remainder of this section, we discuss partial widths of the most relevant decay modes of the additional Higgs bosons. 

The partial decay width for a CP-even mass eigenstate $h_i$ into pairs of vector bosons is given by
\begin{align} \label{eq:width_ZZ}
	\Gamma(h_i \to Z Z) &= \frac{\left(S_{h_i}^{\rm SM}\right)^2 m_Z^4}{16\pi m_{h_i} v^2} \left( 3 - \frac{m_{h_i}^2}{m_Z^2} + \frac{m_{h_i}^4}{4 m_Z^4} \right) \sqrt{1-4 \frac{m_Z^2}{m_{h_i}^2}} \;,
	\\ \Gamma(h_i \to W^+ W^-) &= \frac{\left( S_{h_i}^{\rm SM}\right)^2 m_W^4}{8 \pi m_{h_i} v^2} \left( 3 - \frac{m_{h_i}^2}{m_W^2} + \frac{m_{h_i}^4}{4 m_W^4} \right) \sqrt{1-4 \frac{m_W^2}{m_{h_i}^2}} \;, \label{eq:width_WW}
\end{align}
whereas such decays are forbidden for CP-odd mass eigenstates at tree level. 

Decays into pairs of SM fermions are allowed for both CP-even and CP-odd Higgs bosons. The corresponding partial widths can be written 
\begin{equation} \begin{split} \label{eq:width_ff}
	\Gamma(\Phi_i \to f\bar{f}) &= \frac{N_c^f}{16\pi} \frac{m_f^2}{v^2} m_{\Phi} \left( 1 - 4 \frac{m_f^2}{m_{\Phi_i}^2} \right)^\gamma \\ 
	& \qquad \times \begin{cases} \left( C_\Phi^{\rm SM} - C_\Phi^{\rm NSM}/\tan\beta \right)^2 ~&, {\rm ~for~up-type~quarks	~}f\;, 
 \\ \left( C_\Phi^{\rm SM} + C_\Phi^{\rm NSM} \times \tan\beta \right)^2 ~&, {\rm ~for~down-type~quarks~and~leptons~}f\;, \end{cases}
\end{split} \end{equation} 
where $N_c^f=3$ ($N_c^f=1$) for decays into SM quarks (leptons), $\gamma=3/2$ ($\gamma = 1/2$), and $C_\Phi^{\rm SM} = S_{h_i}^{\rm SM}$ ($C_\Phi^{\rm SM} = 0$) for $\Phi$ CP-even~(CP-odd). Here and in the following, we use $C_\Phi^J$ to refer to the mixing angles for both the CP-even and the CP-odd mass eigenstates $\Phi$ with mass $m_\Phi$; for example, $C_\Phi^{\rm NSM} = S_{h_i}^{\rm NSM}$ or $P_{a_i}^{\rm NSM}$ depending on context. 

If there are additional Majorana fermions $\chi_i$, e.g. Dark Matter candidates, which couple to the Higgs bosons via a coupling $g_{\Phi_i\chi_j\chi_k}$~\footnote{Interactions with the doublet Higgs fields may be generated via their mixing with the singlet or from higher dimensional operators as would result from integrating out a heavy Dirac fermion doublet~\cite{Baum:2017enm}.}
\begin{equation}\label{eq:int_chichi}
	\mathcal{L} \supset \frac{g_{h_i\chi_j\chi_k}}{2(1 + \delta_{jk})} h_i \bar{\chi}_j \chi_k + \frac{g_{a_i\chi_j\chi_k}}{2(1 + \delta_{jk})} a_i \bar{\chi}_j \gamma_5 \chi_k \;,
\end{equation}
the corresponding partial width is given by
\begin{equation} \label{eq:width_chichi}
	\Gamma(\Phi_i \to \chi_j \chi_k) = \left( \frac{2}{1+\delta_{ij}} \right) \frac{g_{\Phi_i\chi_j\chi_k}^2}{16\pi} m_{\Phi_i} \left[ 1- \frac{\left(m_{\chi_j}+m_{\chi_k}\right)^2}{m_{\Phi_i}^2} \right]^{(1+\gamma)} \left[ 1- \frac{\left(m_{\chi_j}-m_{\chi_k}\right)^2}{m_{\Phi_i}^2} \right]^{(1-\gamma)} \;,
\end{equation}
where $\gamma=1/2$ ($\gamma=-1/2$) for a CP-even (CP-odd) $\Phi_i$.

In addition to the above, the extended Higgs sector of the 2HDM+S model allows potentially large decay widths of the heaviest Higgs bosons into pairs of lighter Higgs bosons or a Higgs and a $Z$ boson. CP-conservation allows decays into pairs of Higgs bosons only if they are of the type ($h_i \to h_j h_k$), ($h_i \to a_j a_k$), or ($a_i \to h_j a_k$), while decays into a $Z$ and a Higgs boson must be of the type ($h_i \to Z a_j$) or ($a_i \to Z h_j$). 
The corresponding partial widths are given by 
\begin{align} \label{eq:width_hh}
   \Gamma\left( \Phi_i \to \Phi_j \Phi_k \right) &= \frac{g_{\Phi_i \Phi_j \Phi_k}^2}{16 \pi m_{\Phi_i}} \left(\frac{1}{1+\delta_{jk}}\right) \sqrt{1- 2\frac{m_{\Phi_j}^2 + m_{\Phi_k}^2}{m_{\Phi_i}^2} + \frac{\left(m_{\Phi_j}^2 - m_{\Phi_k}^2\right)^2}{m_{\Phi_i}^4}} \;, \\
	\Gamma(\Phi_i \to Z \Phi_j) &= \frac{\left(C_{\Phi_i}^{\rm NSM} C_{\Phi_j}^{\rm NSM}\right)^2}{32 \pi} \frac{m_Z^2}{m_{\Phi_i} v^2 } \left[ \frac{ \left( m_{\Phi_i}^2 - m_{\Phi_j}^2\right)^2 }{m_Z^2} - 2\left(m_{\Phi_i}^2 + m_{\Phi_j}^2 \right) + m_Z^2 \right] \nonumber
	\\ & \qquad \times \sqrt{1 - 2 \frac{m_{\Phi_j}^2 + m_Z^2}{m_{\Phi_i}^2} + \frac{\left( m_{\Phi_j}^2 - m_Z^2 \right)^2}{m_{\Phi_i}^4}} \;, \label{eq:width_Zh}
\end{align}
where the trilinear couplings between the Higgs mass eigenstates are given in Eq.~\eqref{eq:tricoup_mass}.

\subsection{Branching Ratios: SM Fermions, Misalignment and the 2HDM Limit} \label{sec:BR}

Considering the decays of the heavy Higgs into pairs of SM particles, it is worth repeating that in the alignment limit only the SM-like mass eigenstate $h_{125}$ decays into pairs of vector bosons, while the corresponding branching ratio for the non SM-like mass eigenstates such as ($H \to ZZ$) vanish. The dominant decay mode into SM particles will thus be into the pair of kinematically accessible SM fermions $m_\Phi > 2m_f$ with the largest $\tan\beta$ suppressed (enhanced) Yukawa coupling
\begin{equation}\label{eq:h_yuk}
	y_{\rm max}^\Phi \equiv \max\limits_{m_\Phi > 2 m_f} (N_c^f \frac{m_f}{v} \tan^\gamma\beta)\;,
\end{equation} 
where $\gamma=-1$ ($\gamma=+1$) for up-type (down-type) fermions accounts for the $\tan\beta$ suppression (enhancement) of the Yukawa couplings. For moderate values of $\tan\beta$, the dominant decay mode will thus be into pairs of top quarks if $m_\Phi > 2 m_t$. However, decays into bottom quarks will dominate over those into top quarks even for $m_\Phi > 2 m_t$ if $\tan\beta \gtrsim \sqrt{m_t/m_b} \approx 6.4$. In the following, we identify the regions of parameter space where other decay modes may dominate over decays into pairs of SM fermions.

First, we note that if kinematically accessible, any of the Higgs bosons can decay into possible additional singlet fermions $\chi_i \chi_j$. Ignoring kinematic factors, such decays would compete with decays into SM fermions if 
\begin{equation}
   \chi_i\chi_j\gtrsim f\bar{f}\;:\quad g_{\Phi_i\chi_j\chi_k} \gtrsim |C_{\Phi_i}^{\rm NSM}| y_{\rm max}^{\Phi_i}, 
\end{equation}
where $g_{\Phi_i\chi_j\chi_k}$ is the coupling of $\Phi_i$ to $\chi_i\chi_j$ as defined in Eq.~(\ref{eq:int_chichi}).

Departures from perfect alignment allow for decays of the non SM-like CP-even states $H$ and $h$ into pairs of vector bosons. From Eqs.~(\ref{eq:width_ZZ}) and (\ref{eq:width_ff}), we see that for $m_{h_i}^2 \gg 4 m_Z^2$, decays into pairs of $Z$ ($W$) bosons will compete with decays into pairs of SM fermions if 
\begin{eqnarray}
   ZZ \gtrsim f\bar{f}\;&:&\quad |S_{h_i}^{\rm SM}| m_{h_i}/2v \gtrsim |S_{h_i}^{\rm NSM}| y_{\rm max}^{h_i} \;, \nonumber \\
   W^+W^-\gtrsim f\bar{f}\;&:&\quad |S_{h_i}^{\rm SM}| m_{h_i}/\sqrt{2}v \gtrsim |S_{h_i}^{\rm NSM}| y_{\rm max}^{h_i}\;,
\end{eqnarray} 
where we have ignored the additional contribution to the effective Yukawa coupling from the SM component of $h_i$. Since $H$ and $h$ are only allowed to have a small $S_{h_i}^{\rm SM}$ component from $h_{125}$ phenomenology, these modes are expected to be very suppressed unless $\tan\beta$ is moderate, rendering both the bottom and top Yukawa couplings small, and $m_{h_i} \gg 2 v$, which would significantly reduce the production cross section of such a Higgs boson at the LHC. 

If kinematically accessible, we find from Eqs.~\eqref{eq:width_Zh} and~\eqref{eq:width_ff}, that  decays of a non SM-like Higgs boson $\Phi_i$ into a $Z$ and a lighter Higgs state $\Phi_j$ will compete with decays into pairs of SM fermions if 
\begin{equation}
   \Phi_j Z\gtrsim f\bar{f}\;:\quad |C_{\Phi_j}^{\rm NSM}| m_{\Phi_i}/2 v \gtrsim y_{\rm max}^{\Phi_i}\;. 
\end{equation}
Here, $C_{\Phi_j}^{\rm NSM}$ is the $S_{h_j}^{\rm NSM}$ ($P_{a_j}^{\rm NSM}$) component of the daughter CP-even (CP-odd) $\Phi_j$. Similar to decays into vector bosons, decays of the type $(A \to Z h_{125})$ are suppressed by the required approximate alignment $S_{h_{125}}^{\rm NSM} \ll 1$, making this mode experimentally challenging. On the other hand, decays of the type $(H \to Z a),~(A \to Z h)$ are not suppressed if kinematically accessible, since $C_{\Phi_j}^{\rm NSM}$ is not constrained by the required alignment of $h_{125}$. However, if simultaneously $C_{\Phi_i}^{\rm S}$ becomes large the production cross section of $\Phi_i$ is suppressed.

Likewise, far from the kinematic edge, from Eqs.~\eqref{eq:width_hh} and~\eqref{eq:width_ff} we find that cascade decays into pairs of lighter Higgs bosons ($\Phi_i \to \Phi_j \Phi_k$) will compete with decays into pairs of SM fermions if 
\begin{equation}
   \Phi_j \Phi_k\gtrsim f\bar{f}\;:\quad g_{\Phi_i \Phi_j\Phi_k}/m_{\Phi_i} \gtrsim |C_{\Phi_i}^{\rm NSM}| y_{\rm max}^{\Phi_i}\;. 
\end{equation}
Here, $g_{\Phi_i \Phi_j \Phi_k}$ is the trilinear coupling between the mass eigenstates participating in the cascade decay, and $C_{\Phi_i}^{\rm NSM}$ is the $S_{h_i}^{\rm NSM}$ ($P_{a_i}^{\rm NSM}$) component of the parent CP-even (CP-odd) $\Phi_i$.

Recalling the form of the trilinear couplings in Sec.~\ref{sec:2HDMS_tricoup}, we observe that decays of the non SM-like Higgs bosons into pairs of SM-like Higgs bosons are suppressed in proximity to the alignment limit, $\{{S_{h_{125}}^{\rm NSM}, S_{h_{125}}^{\rm S}} \} \to 0$. For example ($H \to h_{125} h_{125}$), to leading order,  is governed by $\left[g_{H h_{125} h_{125}}\sim 2 S_{H}^{\rm NSM}\left( S_{h_{125}}^{\rm NSM} g_{H^{\rm SM} H^{\rm NSM} H^{\rm NSM}} +  S_{h_{125}}^{\rm S} g_{H^{\rm SM} H^{\rm NSM} H^{\rm S} }\right)\right]$. Thus, for decays into pairs of $h_{125}$'s to compete with decays into $f\bar{f}$, 
\begin{equation}
   h_{125} h_{125} \gtrsim f\bar{f}\;:\quad 2 \left( S_{h_{125}}^{\rm NSM} g_{H^{\rm SM} H^{\rm NSM} H^{\rm NSM}} +  S_{h_{125}}^{\rm S} g_{H^{\rm SM} H^{\rm NSM} H^{\rm S} }\right) /m_H \gtrsim y_{\rm max}^{\Phi_i}\;.
\end{equation}
From the form of the trilinear couplings tabulated in Tab.~\ref{tab:2HDMStricoup}, we see that $\left[g_{H^{\rm SM} H^{\rm NSM} H^{\rm NSM}}\sim \mathcal{O}(v)\right]$, and  $\left[g_{H^{\rm SM} H^{\rm NSM} H^{\rm S}} = \mathcal{M}^2_{S,23}/(\sqrt{2} v) = S_H^{\rm NSM} S_H^{\rm S} (m_H^2-m_h^2)/(\sqrt{2} v)\right]$. Hence, for $m_H \sim v$ and low to moderate values of $\tan\beta$, $g_{H h_{125} h_{125}}/m_H$ can be of the order of the bottom Yukawa. Above the top threshold, $g_{H h_{125} h_{125}}/m_H$ can only compete with the top Yukawa if  $g_{H^{\rm SM} H^{\rm NSM} H^{\rm S}}$ is large enough to compensate for the small $S_{h_{125}}^{\rm S}$, i.e. if $H$ has a substantial singlet component, and if $H$ is significantly heavier than $h$. However, in this case the production cross section of $H$ would be significantly reduced, making this channel challenging. 

The trilinear couplings governing the decays of $(\Phi_i \to h_{125}\Phi_j)$, for $\Phi_j\neq h_{125}$, are not suppressed by alignment. For example, decays of $(H \to h_{125}~ h)$ and $(A \to h_{125}~ a)$ are governed by
\begin{align}
   g_{h_{125} H h} &= \left[ (S_H^{\rm NSM})^2 - (S_H^{\rm S})^2 \right] g_{H^{\rm SM} H^{\rm NSM} H^{\rm S}} + S_H^{\rm NSM} S_H^{\rm S}~\widetilde{g}_H \;,\\
   g_{h_{125} A a} &= \left[ (P_A^{\rm NSM})^2 - (P_A^{\rm S})^2 \right] g_{H^{\rm SM} A^{\rm NSM} A^{\rm S}} + P_A^{\rm NSM} P_A^{\rm S}~\widetilde{g}_A \;.
\end{align}
respectively, where we have defined
\begin{align} \label{eq:gHeff}
 \widetilde{g}_H &\equiv \left( g_{H^{\rm SM} H^{\rm S} H^{\rm S}} - g_{H^{\rm SM} H^{\rm NSM} H^{\rm NSM}} \right) \;,\\
 \widetilde{g}_A &\equiv \left( g_{H^{\rm SM} A^{\rm S} A^{\rm S}} - g_{H^{\rm SM} A^{\rm NSM} A^{\rm NSM}} \right)\;, \label{eq:gAeff}
\end{align}
and assumed perfect alignment, implying $S_H^{\rm NSM} = S_h^{\rm S}$ etc. Since the couplings $g_{H^{\rm SM} H^{\rm NSM} H^{\rm S}}$ and $g_{H^{\rm SM} A^{\rm NSM} A^{\rm S}}$ are proportional to the entries of the mass matrices, these couplings can be rewritten as
\begin{align}\label{eq:freecoup_combination}
   g_{h_{125} H h} &= \frac{S_H^{\rm NSM} S_H^{\rm S}}{\sqrt{2} v} \left\{ \left[ 1 - 2(S_H^{\rm S})^2 \right] \left( m_H^2 - m_h^2 \right) + \sqrt{2} v~\widetilde{g}_H \right\} \;,\\
   g_{h_{125} A a} &= \frac{P_A^{\rm NSM} P_A^{\rm S}}{\sqrt{2}v} \left\{ \left[ 1 - 2(P_A^{\rm S})^2 \right] \left( m_A^2 - m_a^2 \right) + \sqrt{2} v~\widetilde{g}_A \right\} \; .
\end{align}
Hence, such decays will compete with $f\bar{f}$ if
\begin{equation}
   h_{125} \Phi_j \gtrsim f\bar{f}\;:\quad \frac{C_{\Phi_i}^{\rm S}}{\sqrt{2}v m_{\Phi_i} } \left\{ \left[ 1 - 2(C_{\Phi_i}^{\rm S})^2 \right] \left( m_{\Phi_i}^2 - m_{\Phi_j}^2 \right) + \sqrt{2} v~\widetilde{g}_{\Phi_i} \right\} \gtrsim y_{\rm max}^{\Phi_i}\;.
\end{equation}
Since there is no additional suppression of this due to  either the NSM or the S component of $h_{125}$, it can be achieved much more easily, even for small values of $C_{\Phi_i}^{\rm S}$ and $m_{\Phi_i}\sim$ 500 GeV. 

The trilinear couplings relevant for decays of the heavy Higgs into daughter Higgs bosons without $h_{125}$, i.e. ($H \to hh$), ($H \to aa$), and ($A \to ha$), are not necessarily suppressed by small mixing angles, and the corresponding branching ratios can be large. However, these branching ratios are mostly governed by  parameters in the scalar potential which are not related to either the mixing angles or physical masses. Hence, these branching ratios can essentially be taken as free parameters in a generic 2HDM+S.

Observe that while ($\Phi_i \to \Phi_j \Phi_k$) cascade decays are controlled by the trilinear couplings $g_{\Phi_i \Phi_j \Phi_k}$ which are a function of the mixing angles, masses, and the Higgs basis trilinear couplings between the Higgs basis states listed in Eq.~\eqref{trilin}, ($\Phi_i \to Z \Phi_j$) cascade decays are solely controlled by the mass spectrum and the mixing angles.
To summarize, we stress that decays of the non SM-like Higgs bosons into pairs of SM-like Higgs bosons, e.g. ($H \to h_{125} h_{125}$) are suppressed in proximity to the alignment limit. Likewise, decays of the CP-odd mass eigenstates into a $Z$ and a SM-like Higgs boson, e.g. ($A \to Z h_{125}$), are suppressed by alignment. On the other hand, decays of the type 
\begin{eqnarray}
   (H \to h_{125} h)&,& (A \to h_{125} a)\;, \\
   (H \to Z a)&,& (A \to Z h)\;,
\end{eqnarray}
are not suppressed, and if kinematically accessible may be the most relevant cascade decays for probing such models at the LHC. This is because if compared to decays into pairs of new light Higgs boson, such as ($H \to hh$) or ($A \to ha$), they offer a final state with known mass and branching ratios into pairs of SM particles, which can be employed to tag such signatures.

Finally, we would like to comment on the limit where the singlet states decouple from the doublet-like Higgs states. Such a situation is achieved if the mixing angles satisfy
\begin{equation} \label{eq:sdecoupl}
   \left(\{S_{h_{125}}^{\rm S}, S_h^{\rm S}, P_a^{\rm S}\} \to 0\;{ \rm and}\;\{S_H^{\rm S}, P_A^{\rm S}\} \to 1\right) \quad{\rm or}\quad \left(\{S_{h_{125}}^{\rm S}, S_H^{\rm S}, P_A^{\rm S}\} \to 0\;{\rm and}\;\{S_h^{\rm S}, P_a^{\rm S}\} \to 1\right) \;,
\end{equation}
where the first (second) option corresponds to the heavier (lighter) non SM-like states $H$ and $A$ ($h$ and $a$) to be comprised of the singlets $H^{\rm S}$ and $A^{\rm S}$. The first option corresponds to the 2HDM limit of the 2HDM+S.

The latter case can be achieved by a conspiracy of parameters yielding $\{\mathcal{M}_{S,23}^2, \mathcal{M}_{P,12}^2\} \to 0$ in addition to the second alignment condition implying $\mathcal{M}_{S,13}^2 \to 0$. Due to the vanishing mixing of the singlet-like and doublet-like states, the singlet-like states do not couple to SM particles and hence can neither be directly produced at the LHC nor decay, since all couplings to other particles vanish. Recall that in the alignment limit the $g_{H^{\rm SM} H^{\rm SM} H^S}$ coupling vanishes as well such that even ($h \to h_{125} h_{125}$) decays are forbidden.  Thus, the singlet-like states $h$ and $a$ can play the role of DM candidates~\cite{Cai:2013zga,Gaitan:2014vfa,Drozd:2014yla,Han:2017etg,Han:2018bni}. If kinematically allowed, the singlets may be produced at the LHC in the decays of $h_{125}$. The couplings corresponding to ($h_{125} \to h h$) and ($h_{125} \to a a$) decays are $g_{H^{\rm SM} H^{\rm S} H^{\rm S}}$ and $g_{H^{\rm SM} A^{\rm S} A^{\rm S}}$ to leading order, respectively. These couplings are unsuppressed, and hence these decays are only constrained by the invisible decay width of $h_{125}$~\cite{Sirunyan:2017exp}. If at least one of the doublet-like states $H$ and $A$ is sufficiently light such that it can readily be produced at the LHC,
\begin{equation}
   (gg \to A \to a h)\;,\quad (gg \to A \to a h_{125})\;, \quad (gg \to H \to h h)\;,\quad (gg \to H \to h h_{125})\;,
\end{equation}
are the only other production channels of the singlet-like states at the LHC. 

In addition, if the doublet-like states $A$ and $H$ have sufficiently different masses and are light enough such that the heavier state is readily produced at the LHC, depending on the mass ordering, one of the Higgs cascade channels
\begin{equation}
   (gg \to A \to Z H) \quad{\rm or   }\quad (gg \to H \to Z A)
\end{equation}
dominates and is expected to be significant because the corresponding branching ratio is maximal for $S_H^{\rm NSM} S_A^{\rm NSM} \to 1$, cf. Eq.~\eqref{eq:width_Zh}. Observe that in this case, the lighter NSM state would also be directly produced at the LHC, and its presence would be probed by standard heavy Higgs searches. 

The first case of decoupling the singlets from the doublets in Eq.~\eqref{eq:sdecoupl}, where the singlets are heavier than the doublets, is readily achieved by
\begin{equation}
   \mathcal{M}_{S,33}^2 \gg \{ \mathcal{M}_{S,11}^2, \mathcal{M}_{S,22} \}^2\quad{\rm and}\quad \mathcal{M}_{P,22}^2 \gg \mathcal{M}_{P,11}^2 \;,
\end{equation}
without requiring the parameters of the model to conspire to suppress the entries of the mass matrix corresponding to singlet-doublet mixing. In this case, if the doublet-like states $h$ and $a$ are sufficiently split in mass and the heavier state light enough to be readily produced at the LHC, the channels
\begin{equation}
   (gg \to a \to Z h) \quad{\rm or} \quad  (gg \to h \to Z a)
\end{equation}
remain effective, cf. the discussion in the previous paragraph. All other Higgs cascades modes are kinematically forbidden, suppressed by alignment, or suppressed by the vanishing production cross section of the singlet-like states at the LHC.

\subsection{Comparison of Higgs Cascades} \label{sec:cascaderatios}

In the previous section we compared the decay widths corresponding to the Higgs cascades to the decays into pairs of SM particles. In the following, we will compare the final state cross sections for the different Higgs cascades to each other, i.e. we identify which Higgs cascade mode is the most relevant for different regions of the 2HDM+S parameter space. In particular, we want to compare cascade decays into lighter Higgs bosons to cascade decays into a $Z$ and a light Higgs boson. The ratio of the production cross sections is given by
\begin{equation} \label{eq:ratio_general}
	\frac{\sigma(gg \to \Phi_i \to \Phi_j \Phi_k)}{\sigma(gg \to \Phi_l \to Z \Phi_m)} = \frac{\sigma(gg \to \Phi_i)}{\sigma(gg \to \Phi_l)} \times \frac{{\rm BR}(\Phi_i \to \Phi_j \Phi_k)}{{\rm BR}(\Phi_l \to Z \Phi_m)} \;,
\end{equation}
in the narrow width approximation, assuming that the production cross section of the heavy Higgs bosons is dominated by gluon fusion. 

In the rest of our analyses and results, for simplicity, we will assume perfect alignment for $h_{125}$. While some misalignment is allowed by current experimental observations, $h_{125}$ can generically only be contaminated by a NSM fraction of $\lesssim$ few \% and a singlet fraction of $\lesssim$ 20\%~(see e.g. Fig.~1 in Ref.~\cite{Carena:2015moc}). In turn, this contamination is precisely the SM component the other Higgs bosons in the model can have. Since we have so far presented all results in terms of mixing angles and masses, the effect of misalignment on the various cross sections can be deduced by assuming a small SM component for the additional Higgs bosons. From Eqs.~\eqref{eq:ggf}--\eqref{eq:h_yuk} it is easy to see that although the allowed misalignment quantitatively affects the final state cross sections, the qualitative behavior of the model will not be affected. We stress that the phenomenology of the extended Higgs sector of the 2HDM+S will thus primarily be dictated by the results we present in the alignment limit.

We first compare the relevance of the two Higgs cascades for the same parent state $\Phi_i = \Phi_l$.
Using Eqs.~\eqref{eq:width_hh} and~\eqref{eq:width_Zh} we find
\begin{align}\label{eq:ratio_HZaS_HhhS}
	\frac{\sigma(gg \to H \to Z a)}{\sigma(gg \to H \to h_{125} h)} &= \left( \frac{P_A^{\rm S}}{S_H^{\rm S}} \right)^2 \frac{ \left( m_H^2 - m_a^2 \right)^2 - 2\left( m_H^2 + m_a^2 \right) m_Z^2 + m_Z^4}{ \left\{ \left[ 1 - 2(S_H^{\rm S})^2 \right] \left( m_H^2 - m_h^2 \right) + \sqrt{2} v ~\widetilde{g}_H\right\}^2 } \nonumber\\
		&\qquad \times \sqrt{ \frac{ 1 - 2 \left(m_a^2 + m_Z^2\right)/m_H^2 + \left(m_a^2 - m_Z^2\right)^2/m_H^4 }{ 1 - 2 \left(m_h^2 + m_{h_{125}}^2\right)/m_H^2 + \left(m_h^2 - m_{h_{125}}^2\right)^2/m_H^4 } } \;, 
\end{align}
\begin{align} \label{eq:ratio_AZhS_AhaS}
	\frac{\sigma(gg \to A \to Z h)}{\sigma(gg \to A \to h_{125} a)} &= \left(\frac{S_H^{\rm S}}{P_A^{\rm S}}\right)^2 \frac{ \left( m_A^2 - m_h^2\right)^2 - 2\left( m_A^2 + m_h^2 \right) m_Z^2 + m_Z^4}{\left\{ \left[ 1 - 2(P_A^{\rm S})^2 \right] \left(m_A^2 - m_a^2 \right) + \sqrt{2} v ~\widetilde{g}_A\right\}^2} \nonumber\\
		&\qquad \times \sqrt{ \frac{ 1 - 2 \left(m_h^2 + m_Z^2\right)/m_A^2 + \left( m_h^2 - m_Z^2 \right)^2/m_A^4 }{ 1- 2 \left(m_a^2 + m_{h_{125}}^2\right)/m_A^2 + \left(m_a^2 - m_{h_{125}}^2\right)^2/m_A^4} } \;,
\end{align}
where we have implicitly assumed perfect alignment, of particular relevance for estimating the trilinear couplings for the mass eigenstates relevant for the denominators in the ratios above~(see discussion in Sec.~\ref{sec:xsec_BR}).  Also observe that for each set of values of the relevant masses and the mixing angle, one can find a value of the relevant combination of Higgs basis couplings $\widetilde{g}_H$ ($\widetilde{g}_A$), defined in Eqs.~\eqref{eq:gHeff},~\eqref{eq:gAeff}, for which the corresponding trilinear coupling between the mass eigenstates $g_{h_{125} H h}$ ($g_{h_{125} A a}$) vanishes exactly, forbidding the decay channels in the denominators. In such a case, the corresponding ratio of Higgs cascade cross sections tends towards infinity. The threshold correction factor in the second line of Eqs.~\eqref{eq:ratio_HZaS_HhhS} and~\eqref{eq:ratio_AZhS_AhaS}, respectively, is approximately equal to unity except close to the kinematic edge, where the sum of the masses of the decay products is approximately equal to the mass of the  parent Higgs boson.

The trilinear couplings $g_{h_{125} H h}$ and $g_{h_{125} A a}$ are controlled by the involved masses, the relevant mixing angle, and one combination of Higgs basis trilinear couplings each. Also accounting for the parameters controlling the cross sections in the numerator of Eqs.~(\ref{eq:ratio_HZaS_HhhS}) and~(\ref{eq:ratio_AZhS_AhaS}), these Higgs cascade ratios are controlled by seven free parameters: The physical masses $\{m_A, m_H, m_{a}, m_{h}\}$, the mixing angles $\{S_H^{\rm S}, P_A^{\rm S}\}$, and one combination of free trilinear couplings, $\widetilde{g}_A$ relevant for ($A \to h_{125} a$) decays or $\widetilde{g}_H$ relevant for ($H \to h_{125} h$) decays. Note that the ratio in Eq.~\eqref{eq:ratio_HZaS_HhhS} is related to that given in Eq.~\eqref{eq:ratio_AZhS_AhaS} by interchanging the relevant masses, mixing angles, and trilinear couplings referring to the CP-even Higgs sector with the corresponding quantities in the CP-odd sector and vice versa. 

The ratios of the final state production cross sections for processes involving different parent Higgs boson are slightly more involved, since one has to take into account the ratio of the gluon fusion production cross sections as well as the ratio of the decay widths. From Eqs.~\eqref{eq:width_hh},~\eqref{eq:width_Zh}, and~\eqref{eq:ratio_general}, we find in the alignment limit
\begin{equation} \begin{split} \label{eq:ratio_AZhS_HhhS}
	& \frac{\sigma(gg \to A \to Z h)}{\sigma(gg \to H \to h_{125} h)} = \frac{\sigma_{ggh}(m_A)}{\sigma_{ggh}(m_H)} \left( \frac{\tau_A f(\tau_A)}{\tau_A - \left(\tau_A-1\right) f(\tau_A)} \right)^2 \left(\frac{ P_A^{\rm NSM}}{S_H^{\rm NSM}}\right)^4 \frac{m_H}{m_A} \frac{\Gamma_H}{\Gamma_A} \\
		& \qquad\qquad \times \frac{ \left( m_A^2 - m_h^2\right)^2 - 2\left( m_A^2 + m_h^2 \right) m_Z^2 + m_Z^4 }{ \left\{ \left[ 1 - 2(S_H^{\rm S})^2 \right] \left( m_H^2 - m_h^2 \right) + \sqrt{2} v~\widetilde{g}_H \right\}^2 } \\
		& \qquad\qquad \times \sqrt{ \frac{ 1 - 2 \left(m_{h}^2 + m_Z^2\right)/m_A^2 + \left( m_{h}^2 - m_Z^2 \right)^2/m_A^4 }{ 1- 2 \left(m_h^2 + m_{h_{125}}^2\right)/m_H^2 + \left(m_h^2 - m_{h_{125}}^2\right)^2/m_H^4 } } \;,
\end{split} \end{equation}
\begin{equation} \begin{split} \label{eq:ratio_HZaS_AhaS}
	&\frac{\sigma(gg \to H \to Z a)}{\sigma(gg \to A \to h_{125} a)} = \frac{\sigma_{ggh}(m_H)}{\sigma_{ggh}(m_A)} \left( \frac{1}{f(\tau_A)} + \frac{\tau_A-1}{\tau_A} \right)^2 \left(\frac{ S_H^{\rm NSM}}{P_A^{\rm NSM}}\right)^4 \frac{m_A}{m_H} \frac{\Gamma_A}{\Gamma_H} \\
		& \qquad\qquad \times \frac{ \left( m_H^2 - m_a^2 \right)^2 - 2\left( m_H^2 + m_a^2 \right) m_Z^2 + m_Z^4 }{ \left\{ \left[ 1 - 2(P_A^{\rm S})^2 \right] \left(m_A^2 - m_a^2 \right) + \sqrt{2} v~\widetilde{g}_A \right\}^2 } \\
		& \qquad\qquad \times \sqrt{ \frac{ 1 - 2 \left(m_a^2 + m_Z^2\right)/m_H^2 + \left( m_a^2 - m_Z^2 \right)^2/m_H^4 }{ 1- 2\left(m_a^2 + m_{h_{125}}^2\right)/m_A^2 + \left(m_a^2 - m_{h_{125}}^2\right)^2/m_A^4 } }\;, 
\end{split} \end{equation}
where $\Gamma_{\Phi_i}$ is the total decay width of the mass eigenstate $\Phi_i$. Similar to the ratios in Eqs.~\eqref{eq:ratio_AZhS_AhaS} and~\eqref{eq:ratio_HZaS_HhhS}, the ratios in Eq.~\eqref{eq:ratio_HZaS_AhaS} and Eq.~\eqref{eq:ratio_AZhS_HhhS} are equivalent under an interchange of all quantities referring to the CP-even Higgs sector with the corresponding quantities in the CP-odd sector, and vice versa, keeping in mind that exchanging the ratio of the gluon fusion production cross sections entails replacing
\begin{equation}
	\left( \frac{\tau_A f(\tau_A)}{\tau_A - \left(\tau_A-1\right) f(\tau_A)} \right)^2 \leftrightarrow \left( \frac{\tau_A f(\tau_A)}{\tau_A - \left(\tau_A-1\right) f(\tau_A)} \right)^{-2} = \left( \frac{1}{f(\tau_A)} + \frac{\tau_A-1}{\tau_A} \right)^2 \;.
\end{equation}

Compared to the ratios in Eqs.~\eqref{eq:ratio_AZhS_AhaS} and~\eqref{eq:ratio_HZaS_HhhS}, Eqs.~\eqref{eq:ratio_AZhS_HhhS} and~\eqref{eq:ratio_HZaS_AhaS} explicitly depend on only three unknown physical masses, since the processes compared share the same non SM-like Higgs boson in the final state, $h$ for Eq.~\eqref{eq:ratio_AZhS_HhhS} and $a$ for Eq.~\eqref{eq:ratio_HZaS_AhaS}. However, Eqs.~\eqref{eq:ratio_AZhS_HhhS} and~\eqref{eq:ratio_HZaS_AhaS} also depend on the ratio of the gluon fusion production cross sections and the total decay widths of the parent heavy Higgs bosons. The total decay widths are well approximated by the summation of the partial widths listed in Eqs.~\eqref{eq:width_ZZ}--\eqref{eq:width_Zh} and in general are functions of the Higgs mass spectrum, the mixing angles, $\tan\beta$, and additional free trilinear couplings between the Higgs bosons. Observe that since these free trilinear couplings can be sufficiently large such that the corresponding decay modes dominate the decay width, in principle they can dramatically affect the ratios in Eqs.~\eqref{eq:ratio_AZhS_HhhS} and~\eqref{eq:ratio_HZaS_AhaS}. However, in certain limits the ratio of the total width can be well approximated rather simply, e.g. if the decay width of both $A$ and $H$ is dominated by the partial width into pairs of top quarks, we can approximate the ratio in the alignment limit for $\{m_A^2, m_H^2\} \gg 4 m_t^2$ as
\begin{equation}
	\frac{\Gamma_H}{\Gamma_A} \approx \frac{m_H}{m_A} \left( \frac{S_H^{\rm NSM}}{P_A^{\rm NSM}}\right)^2 \;.
\end{equation}

\begin{figure}
	\includegraphics[height=200pt,trim={0cm, 0cm, 4.5cm, 0cm},clip]{{{final_AZhS_AhaS_evenMix_oddMix_mA_750_mH_750_maS_300_mhS_300_freecoup_0}}}
	\includegraphics[height=200pt]{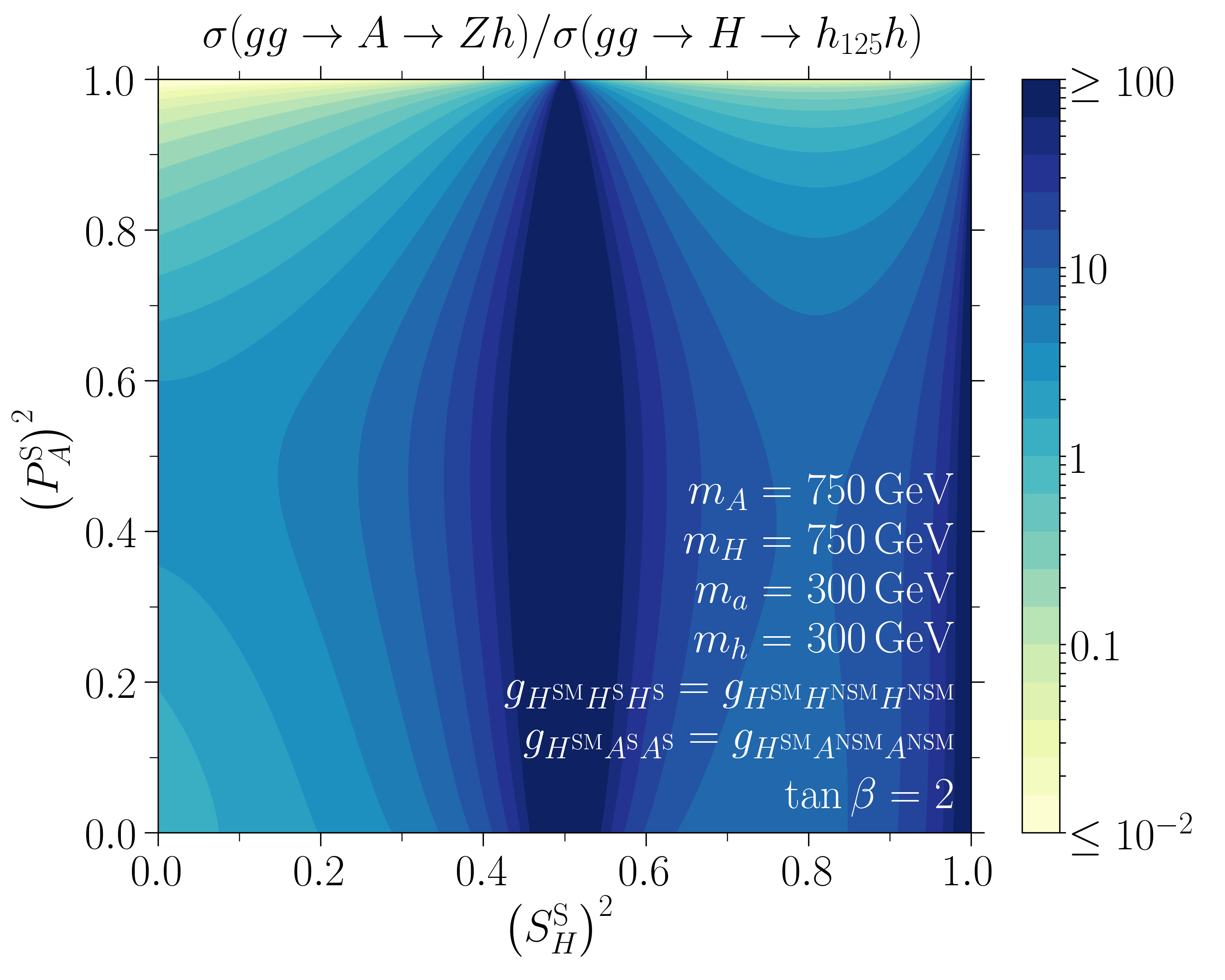}
	\caption{The ratios of the labeled final state cross sections for the Higgs cascades are shown in the plane of the singlet fraction of the parent CP-even and CP-odd heavy Higgs bosons, respectively. {\it Left:} [$\sigma(gg \to A \to Z h) / \sigma(gg \to A \to h_{125} a)$], cf. Eq.~\eqref{eq:ratio_AZhS_AhaS}, {\it Right:} [$\sigma(gg \to A \to Z h) / \sigma(gg \to H \to h_{125} h)$], cf. Eq.~\eqref{eq:ratio_AZhS_HhhS}. The numerical values corresponding to the colors are displayed in the scale legend. Other relevant parameters are fixed to the values indicated in the respective panels. In addition, the remaining trilinear couplings between Higgs basis eigenstates, including the ones which affect the total decay widths of the heavy Higgs bosons $A$ and $H$ relevant for the right panel, are set to 0.}
	\label{fig:ratio_mixings}
\end{figure}

\begin{figure}
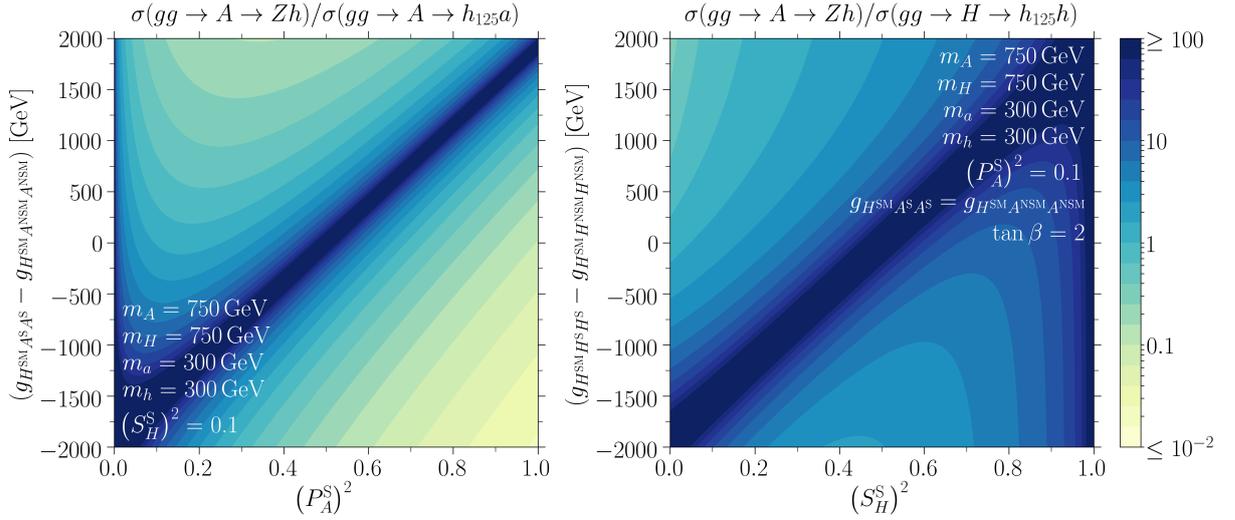

	\includegraphics[height=200pt,trim={0cm, 0cm, 4.5cm, 0cm},clip]{{{final_AZhS_AhaS_freecoup_oddMix_mA_750_mH_750_maS_300_mhS_300_evenMix_0.1}}}
	\includegraphics[height=200pt]{{{final_AZhS_HhhS_freecoupEven_evenMix_mA_750mH_750maS_300_mhS_300_oddMix_0.1_freecoup_odd_0_tanb_2}}}
	\caption{Same as Fig.~\ref{fig:ratio_mixings} but in the plane of the singlet component of the heavy CP-odd (CP-even) Higgs boson $A$ ($H$) for the left (right) panel and the combination of the trilinear couplings $\widetilde{g}_A = (g_{H^{\rm SM} A^{\rm S} A^{\rm S}} - g_{H^{\rm SM} A^{\rm NSM} A^{\rm NSM}})$ (left) and $\widetilde{g}_H = (g_{H^{\rm SM} H^{\rm S} H^{\rm S}} - g_{H^{\rm SM} H^{\rm NSM} H^{\rm NSM}})$ (right) between Higgs basis eigenstates relevant for the respective channel in the denominator of the ratio shown.}
	\label{fig:ratio_freecoup_mix}
\end{figure}

\begin{figure}
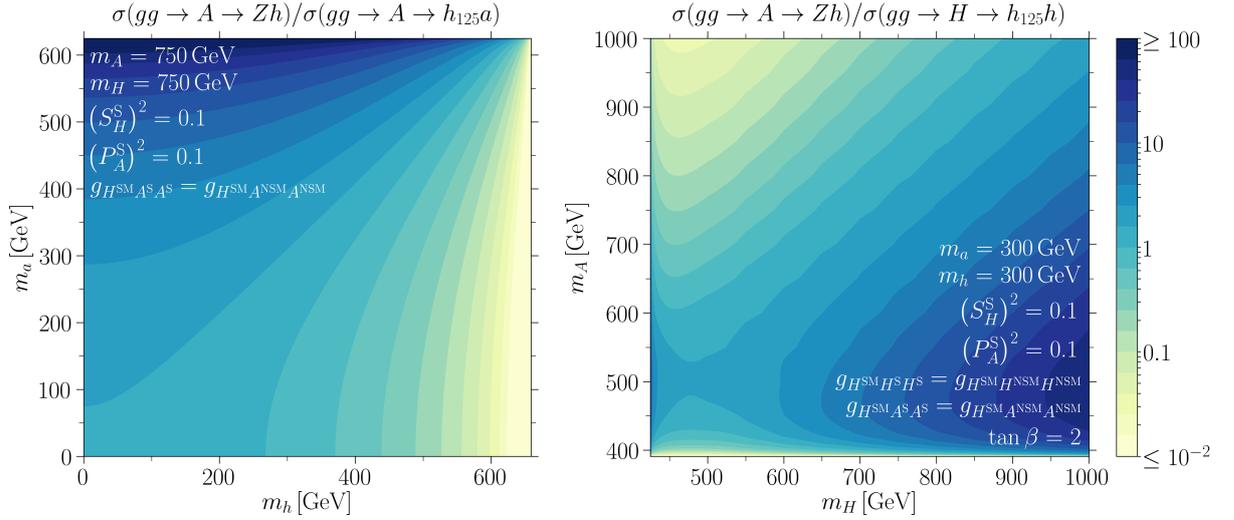

	\includegraphics[height=200pt,trim={0cm, 0cm, 4.5cm, 0cm},clip]{{{final_AZhS_AhaS_maS_mhS_mA_750_mH_750_evenMix_0.1_oddMix_0.1_freecoup_0}}}
	\includegraphics[height=200pt]{{{final_AZhS_HhhS_mA_mH_maS_300_mhS_300_evenMix_0.1_oddMix_0.1_freecoup_even_0_freecoup_odd_0_tanb_2}}}
	\caption{Same as Fig.~\ref{fig:ratio_mixings} but in the plane of the most relevant masses: ($m_h$ vs. $m_a$) in the left panel and ($m_H$ vs $m_A$) in the right panel.}
	\label{fig:ratio_masses}
\end{figure}

In Figs.~\ref{fig:ratio_mixings}--\ref{fig:ratio_masses} we show the ratios of the final state cross sections for the cascade decays in the most relevant parameter planes, fixing the remaining parameters to fiducial values. Note that, as discussed previously, the complimentary ratios not shown can be obtained by swapping all parameters referring to  the CP-even sector with the corresponding quantities in the CP-odd sector. We have chosen the mass scales of the involved particles in the range where the LHC would be most sensitive: heavy Higgs bosons with masses above $\sim 1\,$TeV are challenging because the gluon fusion production cross section drops sharply with mass, while much smaller masses of mostly doublet-like additional Higgs bosons are constrained by direct Higgs searches. 

In Fig.~\ref{fig:ratio_mixings} we show the ratios in the plane of the singlet components of $A$ and $H$. For the particular values of $\widetilde{g}_A=0$~(left) [$\widetilde{g}_H=0$~(right)] chosen here, and for fixed masses, the trilinear couplings $g_{h_{125} A a}$ ($g_{h_{125} H h}$) relevant for the topologies in the denominators are completely controlled by the mixing angles, and are proportional to $C_\Phi^{\rm NSM} C_\Phi^{\rm S} [1-2 (C_\Phi^{\rm S})^2]$, where $\Phi$ is A~(H). As discussed below Eq.~\eqref{eq:freecoup_combination}, these ratios tend towards infinity when the parameters are such that the trilinear couplings vanish. We can see this effect clearly from the fact that both ratios become maximal if the singlet component relevant for the cascade channel in the denominator is approximately $(C_\Phi^{\rm S})^2 = 0.5$. On the other hand, the coupling for both the numerators is controlled by the NSM component of the CP-odd and CP-even states in the topology: $P_A^{\rm NSM} S_h^{\rm NSM} = P_A^{\rm NSM} S_H^{\rm S}$. Hence, the remaining behavior of [$\sigma(gg \to A \to Z h) / \sigma(gg \to A \to h_{125} a)$] shown in the left panel can be understood from observing that this ratio is proportional to $(S_H^{\rm S}/P_A^{\rm S})^2$, cf. Eq.~\eqref{eq:ratio_AZhS_AhaS}. Likewise, the behavior of the ratio [$\sigma(gg \to A \to Z h) / \sigma(gg \to H \to h_{125} h)$] shown in the right panel can be understood from the fact that, ignoring the dependence on the ratio of the total decay widths, but incorporating the effect of the mixing angles on the respective production cross sections, this ratio is proportional to $(P_A^{\rm NSM}/S_H^{\rm NSM})^4 = \left\{ [1-(P_A^{\rm S})^2]/[1-(S_H^{\rm S})^2] \right\}^2$, cf. Eq.~\eqref{eq:ratio_AZhS_HhhS}. 

In Fig.~\ref{fig:ratio_freecoup_mix} we show the same ratios in the planes of the parameters which, apart from the mass spectrum, control the mass eigenstates couplings $g_{h_{125} H h}$ and $g_{h_{125} A a}$ relevant for the decay channel in the denominator: $\widetilde{g}_A$ vs. $(P_A^S)^2$ for the left panel and $\widetilde{g}_H$ vs. $(S_H^S)^2$ for the right panel.  We again see that the ratio in the left (right) panel tends towards infinity when $g_{h_{125} A a}$ ($g_{h_{125} H h}$) is vanishing, i.e. when $\widetilde{g}_A$ ($\widetilde{g}_H$) shown on the y-axis is proportional to [$1 - (C_\Phi^S)^2$]. Away from this region of vanishing trilinear couplings and apart from the dependance of the ratios on the mixing angles discussed in the previous paragraph, from Eqs.~\eqref{eq:ratio_AZhS_AhaS} and~\eqref{eq:ratio_AZhS_HhhS} one finds that for large values of $\widetilde{g}_A$ ($\widetilde{g}_H$) the ratio is inversely proportional to the square of that combination.

Fig.~\ref{fig:ratio_masses} shows the ratios in the plane of the most relevant masses. For the ratio [$\sigma(gg \to A \to Zh)/\sigma(gg \to A \to h_{125} a)$] shown in the left panel and likewise the ratio [$\sigma(gg \to H \to Za)/\sigma(gg \to H \to h_{125} a)$]~(not shown), these are the light non SM-like mass eigenstates in the final states, $h$ and $a$. The behavior shown in the plot is easy to understand: both the numerator and the denominator of the ratio increase in a similar fashion as the difference between the mass of the parent Higgs boson and the sum of the masses of final state particles become larger. On the other hand, these ratios are almost independent of the masses of $H$ and $A$ since the compared topologies are produced via the same parent particle. The right panel displays the ratio [$\sigma(gg \to A \to Zh)/\sigma(gg \to H \to h_{125} h)$] in the plane of the masses of the parent particles $A$ and $H$. The ratio [$\sigma(gg \to H \to Za)/\sigma(gg \to A \to h_{125} a)$]~(not shown) has similar behavior. Both these ratios depend only weakly on the mass of the light non SM-like states $h$ and $a$ since the compared topologies involve the same light Higgs bosons in the final state. However, these ratios depend strongly on the mass of the parent heavy Higgs bosons, $A$ and $H$: The gluon fusion production cross section of a Higgs boson sharply decreases at the LHC with growing mass. This causes the total cross sections of the corresponding cascade topologies to decrease with growing mass of the parent Higgs boson, even though the relevant widths for the decay of the parent state increase with growing mass. In addition, the ratio shown in the right panel of Fig.~\ref{fig:ratio_masses} depends on the ratio of the total decay width of the involved parent Higgs bosons $\Gamma_H/\Gamma_A$. For the range of masses displayed, no strong effects are expected from the total decay widths. However, one finds more pronounced effects for different choices of the masses; in particular threshold effects, e.g. when $m_A$ or $m_H$ falls below $2 m_t \approx 350\,$GeV. Then, the corresponding decay mode, which may have had a large branching ratio, becomes kinematically forbidden. 

In summary, we observe from Figs.~\ref{fig:ratio_mixings}--\ref{fig:ratio_masses} that while there may well be a conspiracy of parameters such that one decay mode is severely suppressed compared to the others, for large parts of the parameter space the cross sections for Higgs cascades with a SM-like Higgs boson and a light additional state $(gg \to \Phi_i \to h_{125} \Phi_j)$ are of the same order of magnitude as the cross sections for the cascades with a $Z$ boson and an additional light Higgs in the final state $(\Phi_i \to Z \Phi_j)$. Hence, one needs to study all possible decay topologies at the LHC in order to comprehensively cover the  2HDM+S parameter space.

\section{Higgs Cascades at the LHC} \label{sec:LHC}

\subsection{Current Limits and Future Projections}
We want to compare the cross section for the cascade decays in the 2HDM+S with the projected reach at the future LHC, assuming a collected luminosity of $L = 3000\,{\rm fb}^{-1}$. In the preceding section we discussed the production of pairs of scalars ($\Phi_j \Phi_k$) or a $Z$ boson and a scalar ($Z \Phi_j$) from an intermediate scalar $\Phi_i$ produced by gluon fusion ($gg \to \Phi_i$). However, at the LHC one would not observe such states directly, but only the decay products of the Higgs and $Z$ bosons. Since the branching ratios into pairs of SM particles as well as the masses of the $Z$ boson and the observed SM-like Higgs state $h_{125}$ are known, they can be used to tag such processes. On the other hand, neither the branching ratios nor the mass of the additional scalars $a$ and $h$ are known. Depending on the mass spectrum and matter content of the model, additional Higgs bosons may decay either in pairs of SM particles ({\it visible}), or, if present and kinematically allowed in pairs of new fermions $\chi$ which, if they are neutral and stable, would leave the detector without depositing energy ({\it invisible}) yielding missing transverse energy (\MET). In the following, we classify the final states arising from such Higgs cascade decays as
\begin{itemize}
	\item $h_{\rm 125}$ + visible,
	\item $h_{\rm 125}$ + invisible, or {\it mono-Higgs},
	\item $Z$ + visible,
	\item $Z$ + invisible, or {\it mono-$Z$},
\end{itemize}
and discuss each signature in detail below.

\subsubsection{$h_{\rm 125}$ + visible:}
Experimental searches in the [$(h_{125} \to b\bar{b}) + (\Phi_j \to q\bar{q}')$] and [$(h_{125} \to b\bar{b}) + (\Phi_j \to b\bar{b})$] final states have been performed by the ATLAS collaboration~\cite{Aaboud:2017ecz, Aaboud:2018knk}. 
The future reach at the LHC in the ($h_{125}$ + visible) channel has been presented in Ref.~\cite{Ellwanger:2017skc} for the $b\bar{b}b\bar{b}$ and $b\bar{b}\gamma\gamma$ final states. 

\subsubsection{ $h_{\rm 125}$ + invisible, or mono-Higgs:}
	The mono-Higgs channel has been discussed in Refs.~\cite{Petrov:2013nia, Carpenter:2013xra,Berlin:2014cfa, No:2015xqa, Basso:2015aee, Abdallah:2016vcn, Liew:2016oon, Bauer:2017ota}. As discussed in Ref.~\cite{Carpenter:2013xra}, the most promising final state is [$(h_{125} \to \gamma\gamma) + \MET$] since the photons in the final state allow for good background discrimination and measurement of \MET. Experimental searches for the mono-Higgs signature have been performed both by the ATLAS and CMS collaborations in a variety of decay modes of $h_{125}$~\cite{Aad:2015yga, Aad:2015dva, Aaboud:2016obm, Sirunyan:2017hnk, Aaboud:2017uak, CMS-PAS-EXO-16-054, Aaboud:2017yqz, CMS-PAS-EXO-16-055}. The future reach at the LHC for the mono-Higgs is presented in Ref.~\cite{Baum:2017gbj}. 

\subsubsection{$Z$ + visible:}
For the ($Z$ + visible) channel, the CMS collaboration has performed a search in the ($Z + b\bar{b}/\tau^+\tau^-$) final states~\cite{Khachatryan:2016are}. No estimate for the future reach at the LHC exists, although the relevance for BSM searches has been pointed out in Refs.~\cite{Carena:2015moc,Baum:2017gbj}.

We extrapolate the reach at the 13\,TeV LHC for 3000\,fb$^{-1}$ of data, $\sigma_{Z+{\rm vis}}^{3000\,{\rm fb}^{-1}}$, from the limits set by the experimental search carried out by the CMS collaboration at $\sqrt{s} = 8\,$TeV with $L = 19.8\,{\rm fb}^{-1}$ of data~\cite{Khachatryan:2016are}. We rescale the reported limit $\sigma_{Z+{\rm vis.}}^{8\,{\rm TeV};\;19.8\,{\rm fb}^{-1}} (m_{\Phi_i}, m_{\Phi_j})$ with the number of events as
\begin{equation}
	\sigma_{Z+{\rm vis.}}^{13\,{\rm TeV};\;3000\,{\rm fb}^{-1}} (m_{\Phi_i}, m_{\Phi_j}) = \sqrt{\frac{\sigma_{ggh}^{8\,{\rm TeV}}(m_{\Phi_i})}{\sigma_{ggh}^{13\,{\rm TeV}}(m_{\Phi_i})} \times \frac{19.8\,{\rm fb}^{-1}}{3000\,{\rm fb}^{-1}}} \times \sigma_{Z+{\rm vis.}}^{8\,{\rm TeV};\;19.8\,{\rm fb}^{-1}} (m_{\Phi_i}, m_{\Phi_j}) \;,
\end{equation}
where $\sigma_{ggh}^{\sqrt{s}}(m)$ is the gluon fusion production cross section of a SM Higgs boson with mass $m$ at the LHC with center-of-mass energy $\sqrt{s}$. Note, that this is a conservative extrapolation of the reach relying purely on the increased statistics, while the ATLAS and CMS collaborations have demonstrated in the past significant improvements on background rejection as well as increased control of the systematic errors when updating searches. 

\subsubsection{$Z$ + invisible, or mono-$Z$:}

\begin{figure}[tbh]
 \includegraphics[width=0.4\linewidth, trim={0cm -6cm 0cm 0cm},clip]{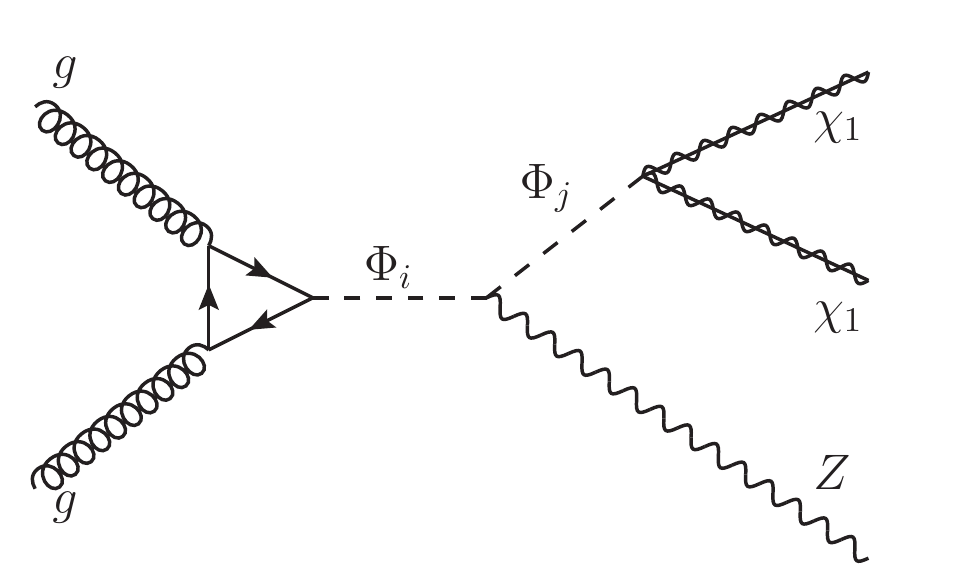}
 \includegraphics[width=0.59\linewidth]{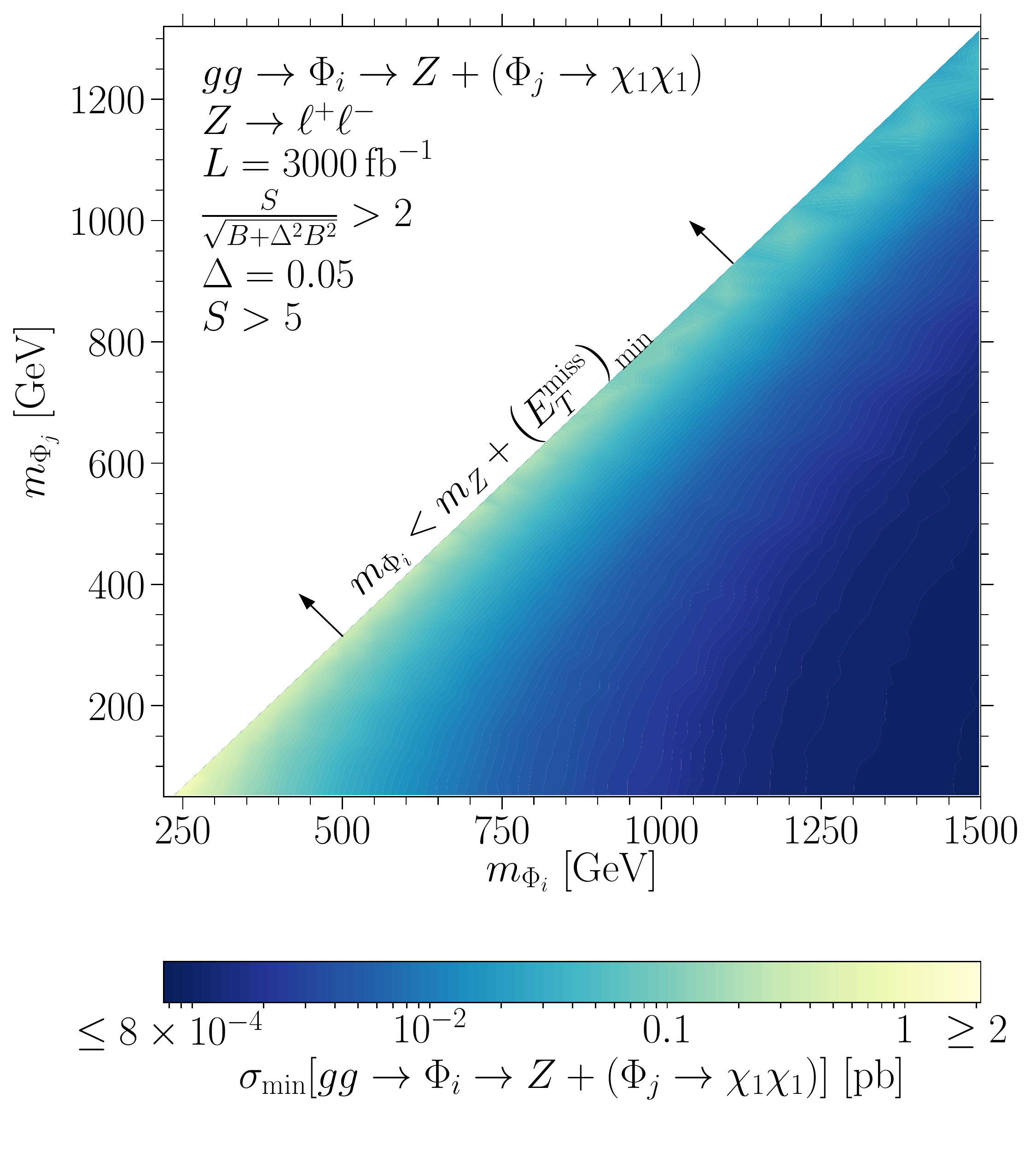}
 \caption{{\it Left:} Illustration of [($gg\to \Phi_i\to Z+(\Phi_j\to\chi_1\chi_1)$] yielding a mono-$Z$ signal at the LHC: $\Phi_i = H$ and $\Phi_j = a$, or $\Phi_i = A$ and $\Phi_j = h$. Decays of the type $A \to Z h_{125}$ are suppressed by alignment, see the discussion in Sec.~\ref{sec:xsec_BR}. {\it Right:} Estimated reach in the scalar topology for a luminosity of $L=3000\,{\rm fb}^{-1}$, assuming a systematic uncertainty of the background of $\Delta = 5\,\%$, shown in the plane of the masses of the involved scalars. The color coding shows the best reach for the ($Z + \MET$) final state, i.e. excluding the $Z \to \ell^+\ell^-$ branching ratio. See Appendix~\ref{app:monoZsim} and text for details.}
	\label{fig:reachScalar}
\end{figure}

Experimental searches for mono-$Z$ have been carried in various final states by both the ATLAS and the CMS collaborations~\cite{Khachatryan:2015bbl,Aaboud:2016qgg,CMS-PAS-EXO-16-038,Aaboud:2017bja,Aaboud:2017rel,Sirunyan:2017hci,Sirunyan:2017jix}. The mono-$Z$ signature has been discussed in Refs.~\cite{Bell:2012rg,Carpenter:2012rg,No:2015xqa,Liew:2016oon,Yang:2017iqh}, although no projection of the future reach at the LHC exists. 

We performed a dedicated collider simulation to estimate the future reach for the mono-$Z$ channel, the details of which are presented in Appendix~\ref{app:monoZsim}. We find that the most relevant topology giving rise to the mono-$Z$ signature proceeds from the decay [$gg\to \Phi_i\to Z+(\Phi_j\to\chi_1\chi_1)$], cf. left panel of Fig.~\ref{fig:reachScalar}. The reach is mainly controlled by the mass splitting $\Delta m_\Phi = \left[ m_{\Phi_1} - \left( m_{\Phi_j} + m_Z \right) \right]$, which controls the \MET\; in the process since the ($\Phi_j \to \chi_1 \chi_1$) system and the $Z$ boson are produced back-to-back from the $s$-channel resonance $\Phi_i$. As long as ($2 m_{\chi_1} < m_{\Phi_j}$), such that the ($\Phi_j \to 2 \chi_1$) decay is kinematically allowed, the reach is virtually independent of the mass of $\chi_1$. Note that it is then straightforward to map the obtained reach to any BSM model containing the required particles by simply computing the corresponding cross section in that model. The reach for this topology from our simulation is presented in Fig.~\ref{fig:reachScalar}. We see that the reach of the mono-$Z$ search at $L = 3000\,{\rm fb}^{-1}$ can be quite powerful. Particularly for $\Delta m_\Phi \gtrsim 300\,$GeV, this channel will allow to probe cross sections as small as $\sigma[gg \to \Phi_i \to Z + (\Phi_j \to \chi_1 \chi_1)] = 10\,$fb.

\subsection{Future Prospects of the 2HDM+S} \label{sec:LHC_2HDMS}
\begin{figure}
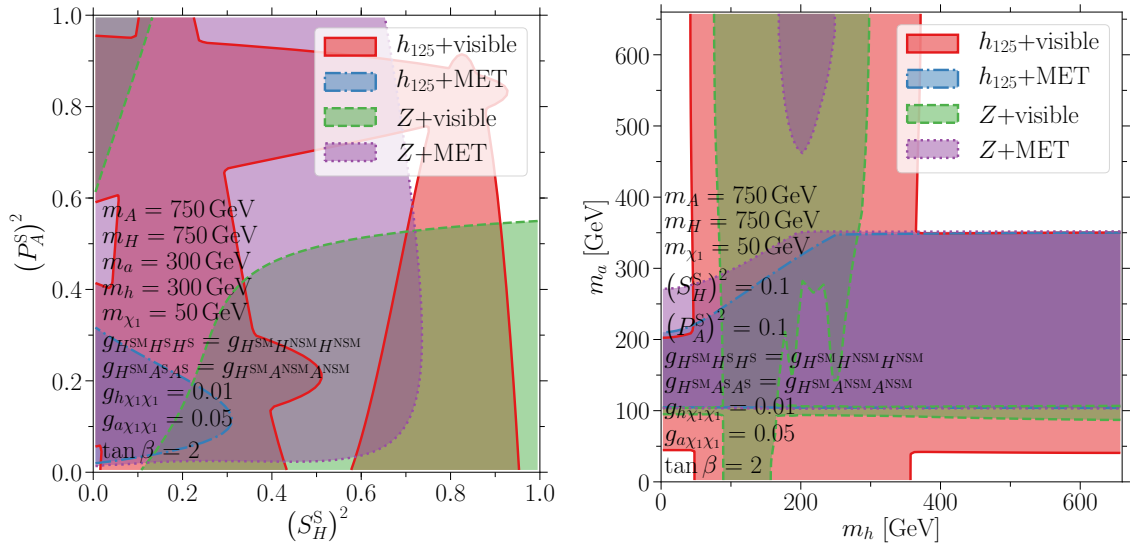

	\includegraphics[width=0.49\linewidth]{{{2HDMS_reach_evenMix_oddMix_mA_750_mH_750_maS_300_mhS_300_mx1_50_freecoupEven_0.0_freecoupOdd_0.0_ghSxx_0.01_gaSxx_0.05_tb_2}}}
	\includegraphics[width=0.49\linewidth]{{{2HDMS_reach_mhS_maS_mA_750_mH_750_mx1_50_evenMix_0.1_oddMix_0.1_freecoupEven_0.0_freecoupOdd_0.0_ghSxx_0.01_gaSxx_0.05_tb_2}}}
	\caption{Regions of 2HDM+S parameter space within the future reach of the different Higgs Cascade search modes as indicated in the legend at the LHC with $L = 3000\,{\rm fb}^{-1}$ of data. The left panel shows the accessible regions in the plane of the singlet fraction of the parent Higgs bosons $(S_H^{\rm S})^2$ vs $(P_A^{\rm S})^2$. The right panel shows the reach in the plane of the masses of the daughter Higgs bosons produced in the Higgs Cascades, $m_h$ vs $m_a$. The remaining parameters are fixed to the values indicated in the labels.}
	\label{fig:2HDMSreach}
\end{figure}

In Sec.~\ref{sec:2HDMS} we compared the production cross sections of the different Higgs cascade decays in the 2HDM+S, in particular, in Sec.~\ref{sec:cascaderatios} we discussed the relations between the 2HDM+S parameters and the prevalence of different Higgs cascade channels. 

In the following, we compute the reach at the future LHC for the cascade topologies in the 2HDM+S discussed above. To this end, we need to include the final state branching ratios of the daughter Higgs bosons produced in the Higgs cascades. For each point in 2HDM+S parameter space, these branching ratios may be obtained from the partial decay widths listed in Sec.~\ref{sec:xsec_BR}. The relevant final states are $\Phi \to \{b\bar{b}, \tau^+ \tau^-, \chi_1\chi_1\}$, where $\Phi = \{h,a\}$ is the mass eigenstate produced as the daughter state in the Higgs cascades. The first two final states are relevant for the ($Z/h_{125}$ + visible) search modes, while the last final state gives rise to mono-Higgs/$Z$ signatures. Note, that the additional fermion $\chi_1$ is not necessarily part of the 2HDM+S model, however if such a fermion were to be stable on collider time scales, its presence would produce the mono-Higgs and mono-$Z$ signatures. If such a fermion is stable on cosmological time scales and is produced in the early Universe via some mechanism, it may yield a viable Dark Matter candidate. This possibility has been discussed in detail in Ref.~\cite{Baum:2017enm}. In general, the couplings of such a fermion to the different 2HDM+S Higgs bosons are interrelated, cf. Ref.~\cite{Baum:2017enm}, however, for the purposes of this work we treat them as free parameters.

In Fig.~\ref{fig:2HDMSreach} we show representative regions of 2HDM+S parameter space within reach of the different Higgs Cascade search modes at the future LHC with 3000\,fb$^{-1}$ of data. We have fixed the parent Higgs boson masses to 750 GeV, high enough that there is no kinematic suppression for any of the SM channels, and standard searches~(in particular $t\bar{t}$ final states) are challenging.\footnote{Note, that the value for the masses of $H/A$ in some figures in this work matches the invariant mass of the (previously observed~\cite{Aaboud:2016tru,Khachatryan:2016hje}, but no longer significant~\cite{Khachatryan:2016yec,Aaboud:2017yyg}) 750\,GeV di-photon excess purely by accident. Since except for the charged Higgs no additional charged particles beyond the SM are present in the model considered here, it would be very challenging, if not impossible, to obtain the required di-photon cross sections to match the previously observed excess.} Other relevant parameters are chosen as shown in the legend. In particular, we choose  $\tan\beta =2$, the mass of the additional fermion is set to $m_{\chi_1} = 50\,$GeV, and its couplings to the light additional Higgs bosons to $g_{h\chi_1\chi_1} = 0.01$ and $g_{a\chi_1\chi_1} = 0.05$. These benchmark values of the couplings are chosen such that as long as decays of the $h/a$ into pairs of top quarks are kinematically forbidden, $g_{h\chi_1\chi_1} \lesssim \{ g_{h f\bar{f}}, g_{a f\bar{f}} \} \lesssim g_{a\chi_1\chi_1}$, where $g_{\Phi f\bar{f}}$ stands for the relevant couplings of $\Phi$ to pairs of SM fermions,
\begin{align}
	g_{\Phi \tau \tau} &= C_\Phi^{\rm NSM} \frac{m_\tau}{v} \tan\beta \approx 0.019 \left(\frac{C_\Phi^{\rm NSM}}{\sqrt{0.9}}\right) \left(\frac{\tan\beta}{2} \right) \;, \\
	g_{\Phi b\bar{b}} &= C_\Phi^{\rm NSM} \frac{m_b}{v} \tan\beta \approx 0.046 \left(\frac{C_\Phi^{\rm NSM}}{\sqrt{0.9}}\right) \left(\frac{\tan\beta}{2} \right) \;,
\end{align} 
in the alignment limit.

The shapes of the regions in Fig.~\ref{fig:2HDMSreach} are determined by a convolution of different factors: the cross section of the different Higgs cascades, the branching ratios in the final states, and the projected reach in the different final state channels. Despite this complication, the general behavior can be understood from what we discussed in Sec.~\ref{sec:cascaderatios} and in particular Figs.~\ref{fig:ratio_mixings}--\ref{fig:ratio_masses}.

We begin by discussing the left panel of Fig.~\ref{fig:2HDMSreach}, where the future reach is shown as a function of the singlet fraction of the parent Higgs bosons $(P_A^{\rm S})^2$ and $(S_H^{\rm S})^2$.  The mass of the daughter Higgs bosons is chosen to be 300 GeV, such that there is no phase space suppression for any of the decays. First, note that the gluon fusion production cross section of $\Phi = \{H,A\}$ is suppressed by the singlet fraction of $\Phi$ since $\sigma(gg \to \Phi) \propto \left[ 1 - (C_\Phi^{\rm S})^2 \right]$, cf. Eq.~\eqref{eq:ggf}, degrading all the corresponding search channels. Due to the chosen values of the couplings of $h$ and $a$ to pairs of the new fermions $\chi_1$, cascade decays of the type ($H \to h_{125} h$) and ($A \to Z h$) will give rise to ($h_{125}$ + visible) and ($Z$ + visible) signatures, respectively, while cascade decays of the type ($A \to h_{125} a$) and ($H \to Z a$) will in general lead to mono-Higgs and mono-$Z$ signals. This behavior is most pronounced when comparing the ($Z$ + visible) with the mono-$Z$ region in the left panel of Fig.~\ref{fig:2HDMSreach}: Large values of $(P_A^{\rm S})^2$ suppress the gluon fusion production cross section of $A$ and hence the cross section of ($A \to Z h$) cascades giving rise to ($Z$ + visible) final states, while large values of $(S_H^{\rm S})$ suppress the gluon fusion production cross section of $H$ and hence ($H \to Z a$) cascades giving rise to mono-$Z$ final states. Note, that since the decay width for ($\Phi_i \to Z \Phi_j$) decays is proportional to $(C_{\Phi_i}^{\rm NSM} C_{\Phi_j}^{\rm NSM})^2$, the cascade decay ($H \to Z a$) [$A \to Z h$] is also suppressed for $(P_A^{\rm S})^{\rm 2} \approx 0$ [$(S_H^{\rm S}) \approx 0$] corresponding to very singlet-like $a$ [$h$], i.e. $C_{\Phi_j}^{\rm NSM} \approx 0$. 

In order to understand the behavior of the mono-Higgs and ($h_{125}$ + visible) regions, recall that for the chosen values of the Higgs basis trilinear couplings ($g_{H^{\rm SM} H^{\rm S} H^{\rm S}} = g_{H^{\rm SM} H^{\rm NSM} H^{\rm NSM}}$)~[($g_{H^{\rm SM} A^{\rm S} A^{\rm S}} = g_{H^{\rm SM} A^{\rm NSM} A^{\rm NSM}}$)], the coupling $g_{h_{125} A a}$ [$g_{h_{125} H h}$] vanishes for $(P_A^{\rm S})^2 \to 0$ [$(S_H^{\rm S})^2 \to 0$], cf. Eqs.~\eqref{eq:ratio_AZhS_AhaS}--\eqref{eq:ratio_HZaS_AhaS} and Fig.~\ref{fig:ratio_mixings}. Since the ($h_{125}$ + visible) decays arise through ($h/a \to b\bar{b}$) decays [while the ($Z$ + visible) reach is obtained from the ($Z \to \tau^+ \tau^-$) final state], the balance between the ($h_{125}$ + visible) and mono-Higgs regions is skewed towards ($h_{125}$ + visible) final states. For larger values of $g_{a\chi_1\chi_1}$ than chosen in Fig.~\ref{fig:2HDMSreach}, one would obtain larger regions where the mono-Higgs final state would be within the future reach of the LHC, while simultaneously decreasing the regions where ($h_{125}$ + visible) final states would be within the LHC reach. 

In the right panel of Fig.~\ref{fig:2HDMSreach} we show the regions within reach of the different Higgs Cascade search modes in the plane of the masses of the Higgs bosons produced in the final state, $m_h$ and $m_a$. The singlet component of the parent Higgs bosons is set to $(C_\Phi^{\rm S})^2=0.1$, implying the the daughter Higgs bosons are dominantly singlet-like, i.e. $\{(S_h^{\rm NSM})^2,(P_a^{\rm NSM})^2\}=0.1$. If decays into pairs of top quarks are kinematically allowed for both $h$ and $a$, i.e. $\{ m_h, m_a \} \gtrsim 350\,$GeV, our 2HDM+S model is not within reach of any of the searches listed since the decays of both $a$ and $h$ are dominated by decays into top quarks. This is because the corresponding coupling $g_{\Phi t\bar{t}} \approx C_\Phi^{\rm NSM} / \tan\beta$ is much larger than the couplings to other SM states or the benchmark values chosen for the $g_{\Phi \chi_1 \chi_1}$ couplings unless $\tan\beta$ is large or $\Phi$ is very singlet like, corresponding to $C_\Phi^{\rm NSM} \to 0$. In this region, the signal would be $t\bar{t}h_{125}$ or $t\bar{t} Z$. Such decays may be probed by SM $t\bar{t} h_{125}$ searches in the near future~\cite{Aaboud:2017rss,Sirunyan:2018hoz}. Observe that this region would be within reach of the mono-$Z$ and mono-Higgs searches for large values of the $g_{\Phi \chi_1 \chi_1} \gtrsim 1$ couplings. Alternatively, if there is some departure from the alignment limit such that $h$ has some non-negligible $H^{\rm SM}$ component, one may search for final states arising from ($h \to ZZ/WW$) decays. 

Note that the cut-off for low masses of $h$ or $a$ for the mono-Higgs and mono-$Z$ searches is due to the ($h/a \to \chi_1 \chi_1$) searches being kinematically forbidden for $m_h/m_a < 2 m_{\chi_1} = 100\,$GeV. For the ($h_{125}/Z$ + visible) final states the cutoff mainly comes from the fact that Refs.~\cite{Ellwanger:2017skc,Khachatryan:2016are} provide the reach only down to a minimal value of $m_h/m_a$. Observe that the region for the ($Z$ + visible) final state is not as smooth as for the other channels, in particular in the region ($150 \lesssim m_h/{\rm GeV} \lesssim 250$, $100 \lesssim m_a/{\rm GeV} \lesssim 300$) because the reach for this final states is extrapolated from the experimental result in Ref.~\cite{Khachatryan:2016are}. Hence, the reach as a function of the involved masses suffers from statistical fluctuations, while the reach in the remaining final states is evaluated from theoretical predictions based on Monte Carlo simulations, which suffer much less from statistical fluctuation and give rise to smooth contours in the mass plane. 

The remaining behavior portrayed in the right panel of Fig.~\ref{fig:2HDMSreach} is determined by what we discussed above for the left panel: for the chosen values of the $g_{\Phi \chi_1\chi_1}$ couplings, mono-Higgs [mono-$Z$] final states are dominantly accessible through ($A \to h_{125} + a$) [$H \to Z a$] Higgs Casacades and hence strongly depend on the mass of $a$. On the other hand, ($h_{125}$ + visible) [$Z$ + visible] final states are dominantly due to ($H \to h_{125} + h$) [$A \to Z + h$] Higgs Casacades and are hence controlled by the mass of $h$. As discussed above, this separation is much clearer when comparing the ($Z$ + visible) and mono-$Z$ signatures than when comparing the ($h_{125}$ + visible) and mono-Higgs channels due to the values chosen for the $g_{\Phi\chi_1\chi_1}$ couplings.

Note that while we have fixed the mass of the parent Higgs bosons at 750 GeV when presenting our results, and assumed perfect alignment, the effect of varying these quantities is easy to deduce. First, different masses for the parent Higgs bosons would primarily affect the gluon fusion production cross section, whose scaling with mass is well known. The affect of misalignment would be quantitatively negligible on the reach of the Higgs Cascades discussed here. However, various decay chains not considered in the above analysis, such as $(H\to h_{125} h_{125})$ or $(A\to Z h_{125})$ would be present. As discussed in Sec.~\ref{sec:BR}, such decays are  suppressed by either the NSM or S component of $h_{125}$ compared to decays into $h$. The reach for such decays can be extrapolated from those presented by convoluting the relevant decay widths with the misalignment  of $h_{125}$ and identifying $m_h =125$ GeV in the right panel of Fig.~\ref{fig:2HDMSreach}.  Observe that our results can also be mapped to the case of the decoupled singlet and non-degenerate CP-odd and CP-even NSM-like Higgs bosons by appropriately choosing the NSM and S components for the parent and daughter Higgs bosons in the decay chain of interest~(c.f. discussion in Sec~\ref{sec:BR}).

Finally, note that Fig.~\ref{fig:2HDMSreach} is meant to portray only the prospects of exploring the 2HDM+S parameter space using Higgs cascades. The regions displayed do not take into account existing bounds from searches for additional Higgs boson beyond $h_{125}$. In particular, the charged Higgs does not play a role in any of the searches shown. Existing constraints on charged Higgs bosons, e.g. from flavor physics observables, can be satisfied by choosing a sufficiently large mass of the charged Higgs. Recall that in this work we treat the masses of the physical Higgs bosons as free parameters. Even without considering effects of mixing, mass splittings of order of a few 100\,GeV between the mostly doublet-like pseudo-scalar and the charged Higgs are easily achievable; from the equations given in Appendix~\ref{app:2HDMS_mass} we find for example [$\mathcal{M}_{P,11}^2 - m_{H^\pm}^2 = v^2 \left( \lambda_4-\lambda_5\right)$], where $\mathcal{M}_{P,11}^2$ is the element of the squared mass matrix corresponding to the $A^{\rm NSM}$ interaction eigenstate. Furthermore, in more complete models with larger particle content than the 2HDM+S considered here such as the NMSSM, indirect observables such as those from flavor physics receive additional contributions beyond those from the charged Higgs which may loosen the bounds on the mass of the charged Higgs, cf. Refs.~\cite{Altmannshofer:2012ks,Carena:2018nlf}.

In summary, as evident from Fig.~\ref{fig:2HDMSreach}, there are large regions of parameter space in reach of the different Higgs Cascade search modes. In particular, the Higgs Cascades enable the LHC to probe regions of parameter space challenging to access with traditional searches for the direct decays of additional Higgs states: Singlet-like light states are difficult to directly produce due to the small couplings to pairs of SM particles. On the other hand, doublet-like states are readily produced, but if their mass is above the kinematic threshold allowing for decays into pairs of top quarks, $\Phi \to t\bar{t}$ decays will dominate over the decays into other SM states. Pairs of top quarks produced from an $s$-channel resonance are very difficult to detect at the LHC due to interference effects with the QCD background, which makes the $m_\Phi \gtrsim 350\,$GeV, low $\tan\beta$ region extremely challenging to probe at the LHC through direct Higgs decays with current search strategies~\cite{Dicus:1994bm,Barcelo:2010bm,Barger:2011pu,Bai:2014fkl,Jung:2015gta,Craig:2015jba,Gori:2016zto,Carena:2016npr}.

\section{Conclusions} \label{sec:conclusions}
The 2HDM+S is well motivated both by phenomenological considerations such as facilitating baryogenesis or DM model building, and since interesting high energy models with such a Higgs sector exist, e.g. the NMSSM. Here, we presented the first systematic study of this model. The seemingly large number of parameters complicates analyzing the model space. By mapping the parameters of the generic Higgs potential to physically relevant quantities like masses and mixing angles, and defining the remaining set of independent trilinear and quartic couplings, we obtain a much more intuitive understanding of the 2HDM+S parameter space. This mapping also makes transparent how embedding the 125\,GeV Higgs boson observed at the LHC and its required SM-like couplings constrain the 2HDM+S parameter space.

The extended Higgs sector of the 2HDM+S allows for a rich phenomenology beyond what is encountered in 2HDMs. In particular, the presence of the singlet allows for interesting cascade decays of the heavy Higgs bosons into two lighter Higgs bosons or a light Higgs and a $Z$ boson. Using such Higgs cascades in addition to the conventional searches for additional Higgs bosons allows to cover a large part of the model space despite the seemingly large number of free parameters. Although short of direct evidence~(such as observing all the 2HDM+S Higgs bosons), Higgs cascades  are also an important handle for differentiating such a model from e.g. the 2HDM, where Higgs cascades are hard to realize.

We compute the various production cross sections and branching ratios of the Higgs bosons (both CP-even and odd), allowing us to analytically understand which model parameters control the phenomenology. We show that cascade decays of a parent Higgs bosons into final states involving a single 125\,GeV Higgs boson or a $Z$ can have relevant cross sections. In particular, we show that while in certain regions of parameter space these two can be complimentary, for the most part, both signatures are comparable. 

For collider searches, one must also consider the decays of the daughter Higgs bosons produced in the Higgs cascade decays. Besides decays into pairs of SM particles, these light Higgs bosons may also decay into new neutral states which, if stable on collider times scales, give rise to missing energy signatures. In order to include models with such additional states, which may serve as a Dark Matter candidate, we add an arbitrary fermion to the 2HDM+S. The final states arising from Higgs cascades can then be categorized as $h_{125}/Z$+visible or $h_{125}/Z$+invisible (mono-Higgs/$Z$). Convoluting the production cross sections and branching ratios and comparing to the estimated sensitivity of the LHC with $L=3000\,{\rm fb}^{-1}$ of data in the various final states, we show that most of the interesting region of 2HDM+S parameter space can be probed by combining the different Higgs cascade modes.

The NMSSM is arguably the best motivated high scale model with a 2HDM+S Higgs sector. The $Z_3$-invariant NMSSM usually discussed in the literature represents a very constrained version of the 2HDM+S with only 7 free parameters controlling the Higgs sector. The general NMSSM is much more complicated, but can be mapped trivially to a generic 2HDM+S. Without any direct evidence at the LHC for non-SM like Higgs bosons, it is important and timely to consider all the different types of signatures that may arise in the generic situation, in particular since the low $\tan\beta$ region of parameter space is challenging to probe experimentally above the top threshold. 

\acknowledgments

We are indebted to Bibhushan Shakya for early collaboration in this project. We would also like to thank Katie Freese and Carlos Wagner for interesting discussions. 
SB would like to thank the LCTP and the University of Michigan as well as Wayne State University, where part of this work was carried out, for hospitality. 
SB acknowledges support by the Vetenskapsr\r{a}det (Swedish Research Council) through contract No. 638-2013-8993 and the Oskar Klein Centre for Cosmoparticle Physics. NRS is supported by DoE grant DE-SC0007983 and Wayne State University. The work of NRS was partially performed at the Aspen Center for Physics, which is supported by National Science Foundation grant PHY-1607611.

\appendix
\newpage
\section{Mass Matrices}\label{app:2HDMS_mass}
The entries of the (symmetric) squared mass matrices are obtained from the scalar potential via
\begin{equation}
	\mathcal{M}_{ij}^2 \equiv \left.\frac{\partial^2 V}{\partial \Phi_i \partial \Phi_j}\right|_{\substack{\Phi_1 = v_1\\\Phi_2 = v_2\\S = v_S}} \;.
\end{equation}
By construction, the neutral and charged Goldstone modes $G^0$ and $G^\pm$, respectively, are massless and do not mix with the remaining states. The mixing of the charged Higgs $H^\pm$ with the neutral states is forbidden by gauge invariance. Finally, CP conservation forbids mixing of CP-even and CP-odd states. Thus, the non-trivial entries of the mass matrix can be split into 6 entries concerning the CP-even Higgs bosons, 3 entries concerning the CP-odd states, and the (squared) mass of the charged Higgs.

The entries of the CP-even Higgs squared mass matrix in the extended Higgs basis $\{H^{\rm SM}, H^{\rm NSM}, H^{\rm S} \}$ at tree level are~\footnote{Recall that we use the interaction basis defined by the Yukawa structure, Eq.~\eqref{eq:Yuk}, and where the singlet field $S$ is shifted such that the tadpole term $\xi = 0$. Please see the Appendices of Ref.~\cite{Carena:2015moc} for the mass matrices in terms of parameters defined in the extended Higgs basis of the 2HDM+S.}
\begin{align} \label{eq:MS11}
	\mathcal{M}_{S,11}^2 &= v^2 \left[ 2 \lambda_1 s_\beta^4 + 2 \lambda_2 c_\beta^4 + \lambda_{345} s_{2\beta}^2 + 4 s_{2\beta} \left(\lambda_6 s_\beta^2 + \lambda_7 c_\beta^2 \right) \right] ,
	\\ \mathcal{M}_{S,12}^2 &= v^2 \left[ s_{2\beta} \left( \lambda_1 s_\beta^2 - \lambda_2 c_\beta^2 \right) + \frac{\lambda_{345}}{2} s_{4\beta} + \left( \lambda_6 + \lambda_7 \right) c_{2\beta} - \left( \lambda_6 - \lambda_7 \right) c_{4\beta} \right],
	\\ \mathcal{M}_{S,13}^2 &= v \left\{ 2 \mu_{11} s_\beta^2 + \mu_{1221} s_{2\beta} + 2 \mu_{22} c_\beta^2 + 2 v_S \left[ \left( \lambda'_1 + 2 \lambda'_4 \right) s_\beta^2 + \lambda'_{367} s_{2\beta} + \left( \lambda'_2 + 2 \lambda'_5 \right) c_\beta^2 \right] \right\},
	\\ \mathcal{M}_{S,22}^2 &= v^2 \left[ \frac{\lambda_1 + \lambda_2}{2} s_{2\beta}^2 - \lambda_{345} s_{2\beta}^2 - \frac{\lambda_6 + \lambda_7}{s_{2\beta}} - \frac{c_{6\beta} - 3 c_{2\beta}}{2 s_{2\beta}} \left(\lambda_6 - \lambda_7\right) \right] \nonumber\\
   & \quad + \frac{2}{s_{2\beta}} \left( m_{12}^2 - v_S \mu_{1221} - v_S^2 \lambda'_{367} \right),
	\\ \mathcal{M}_{S,23}^2 &= v \left\{ \left( \mu_{11} - \mu_{22} \right) s_{2\beta} + \mu_{1221} c_{2\beta} + v_S \left[ \left( \lambda'_1 - \lambda'_2 + 2\lambda'_4 - 2\lambda'_5 \right) s_{2\beta} + 2 \lambda'_{367} c_{2\beta} \right] \right\}, 
	\\ \mathcal{M}_{S,33}^2 &= - \frac{v^2}{2 v_S} \left( 2 \mu_{11} s_\beta^2 + \mu_{1221} s_{2\beta} + 2 \mu_{22} c_\beta^2 \right) + \frac{v_S^2}{3} \left( \lambda''_1 + 4 \lambda''_2 + 3 \lambda''_3 \right) + \frac{v_S}{2} \left( \mu_{S1} + 3 \mu_{S2} \right) ~,
\end{align}
where
\begin{align}
	\lambda_{345} &\equiv \lambda_3 + \lambda_4 + \lambda_5 \;,
	\\ \mu_{1221} &\equiv \mu_{12} + \mu_{21} \;,
	\\ \lambda'_{367} &\equiv \lambda'_3 + \lambda'_6 + \lambda'_7 \;,
\end{align}
and we used a shorthand notation $s_\beta \equiv \sin\beta$, $c_\beta \equiv \cos\beta$, etc.

The entries of the (symmetric) squared mass matrix for the CP-odd Higgs bosons in the basis $\{A^{\rm NSM}, A^{\rm S}\}$ are at tree level given by
\begin{align}
	\mathcal{M}_{P,11}^2 &= -v^2 \left( 2 \lambda_5 + \lambda_6 t_\beta + \frac{\lambda_7}{t_\beta} \right) + \frac{2}{s_{2\beta}} \left( m_{12}^2 - v_S \mu_{1221} - v_S^2 \lambda'_{367} \right) ,
	\\ \mathcal{M}_{P,12}^2 &= v \left[ \mu_{12} - \mu_{21} + 2 v_S \left( \lambda'_6 - \lambda'_7 \right) \right] ,
	\\ \mathcal{M}_{P,22}^2 &= - v^2 \left[ \frac{\mu_{11} s_\beta^2 + \mu_{1221} s_{2\beta}/2 + \mu_{22} c_\beta^2}{v_S} + 4 \lambda'_4 s_\beta^2 + 2 \left(\lambda'_6 + \lambda'_7\right) s_{2\beta} + 4 \lambda'_5 c_\beta^2 \right] + \nonumber
	\\ & \qquad - 2 m_S'^2 - \frac{v_S}{2} \left( 3 \mu_{S1} + \mu_{S2} \right) - \frac{2}{3} v_S^2 \left( \lambda''_1 + \lambda''_2 \right) \;. \label{eq:MP22}
\end{align}
The mass of the charged Higgs boson is given by
\begin{equation}
	m_{H^\pm}^2 = -v^2 \left( \lambda_4 + \lambda_5 + \lambda_6 t_\beta + \frac{\lambda_7}{t_\beta} \right) + \frac{2}{s_{2\beta}} \left( m_{12}^2 - v_S \mu_{1221} - v_S^2 \lambda'_{367} \right) \;.
\end{equation}

\section{Trilinear couplings}\label{app:2HDMS_tricoup}
The trilinear couplings are obtained from the scalar potential via
\begin{equation}
 g_{\Phi_i\Phi_j\Phi_k}^2 \equiv \left.\frac{\partial^3 V}{\partial \Phi_i \partial \Phi_j \partial \Phi_k}\right|_{\substack{\Phi_1 = v_1\\\Phi_2 = v_2\\S = v_S}} \;.
\end{equation}
We list them for the interaction states of the extended Higgs basis in Table~\ref{tab:2HDMStricoup} in terms of the parameters in the scalar potential\footnote{Recall that we use the interaction basis defined by the Yukawa structure, Eq.~\eqref{eq:Yuk}, and where the singlet field $S$ is shifted such that the tadpole term $\xi = 0$. Please see the Appendices of Ref.~\cite{Carena:2015moc} for the trilinear couplings in terms of parameters defined in the extended Higgs basis of the 2HDM+S.}. Couplings which are strictly vanishing are marked green, while couplings which are proportional to other couplings are marked blue. Besides those couplings marked in the tables, the couplings
\begin{align}
	g_{H^{\rm SM} A^{\rm NSM} A^{\rm NSM}} &= g_{H^{\rm SM} H^{\rm NSM} H^{\rm NSM}} - \frac{\sqrt{2}}{v} \left( \mathcal{M}_{S,22}^2 - \mathcal{M}_{P,11}^2 \right) ,\\
	g_{H^{\rm SM} H^+ H^-} &= g_{H^{\rm SM} H^{\rm NSM} H^{\rm NSM}} - \frac{\sqrt{2}}{v} \left( \mathcal{M}_{S,22}^2 - m_{H^\pm}^2 \right) ,
\end{align}
can be written in terms of the entries of the mass matrices.

\begin{table}[h!]
 {\small
 \begin{center}
 \setlength{\extrarowheight}{4.5pt}
 \begin{tabular}{rl}
 \hline\hline
 {\normalsize $\left(\Phi^i \Phi^j \Phi^k\right)$} : & {\normalsize $\sqrt{2} g_{\Phi^i \Phi^j \Phi^k}$}\\
 \hline
 $\left(H^{\rm SM} H^{\rm SM} H^{\rm SM}\right)$ : & $3\mathcal{M}_{S,11}^2/v$ \\
 $\left(H^{\rm SM} H^{\rm SM} H^{\rm NSM}\right)$ : & $3 \mathcal{M}_{S,12}^2/v$ \\
 $\left(H^{\rm SM} H^{\rm SM} H^{\rm S}\right)$ : & $ \mathcal{M}_{S,13}^2/v$ \\
 $\left(H^{\rm SM} H^{\rm NSM} H^{\rm NSM}\right)$ : & $v \left[ 3 \left( \lambda_1 + \lambda_2 \right) s_{2\beta}^2/2 + \left( \lambda_3 + \lambda_4 + \lambda_5 \right) \left( c_{2\beta}^2 + c_{4\beta} \right) + 3 \left(\lambda_6 - \lambda_7\right) s_{4\beta} \right]$ \\
 $\left(H^{\rm SM} H^{\rm NSM} H^{\rm S}\right)$ : & $\mathcal{M}_{S,23}^2/v$ \\
 $\left(H^{\rm SM} H^{\rm S} H^{\rm S}\right)$ : & $2 v \left[ \left( \lambda'_1 + 2 \lambda'_4 \right) s_\beta^2 + \left( \lambda'_3 + \lambda'_6 + \lambda'_7 \right) s_{2\beta} + \left( \lambda'_2 + 2 \lambda'_5 \right) c_\beta^2 \right] $ \\
 $\left(H^{\rm NSM} H^{\rm NSM} H^{\rm NSM}\right)$ : & $3 v s_{2\beta} \left[ \lambda_1 c_\beta^2 - \lambda_2 s_\beta^2 - \left( \lambda_3+\lambda_4+\lambda_5\right) c_{2\beta} + \frac{\lambda_6}{t_\beta} - 2\left(\lambda_6 - \lambda_7\right) s_{2\beta} - \lambda_7 t_\beta \right] $ \\
 $\left(H^{\rm NSM} H^{\rm NSM} H^{\rm S}\right)$ : & $2 \left\{ \mu_{11} c_\beta^2 - \left(\mu_{12}+\mu_{21}\right) s_{2\beta}/2 + \mu_{22} s_\beta^2 + \right.$
 \\ & $ \quad \left.+ v_S \left[ \left( \lambda'_1 + 2\lambda'_4 \right) c_\beta^2 - \left(\lambda'_3 + \lambda'_6 + \lambda'_7\right) s_{2\beta} + \left( \lambda'_2 + 2\lambda'_5 \right) s_\beta^2 \right] \right\}$ \\
 $\left(H^{\rm NSM} H^{\rm S} H^{\rm S}\right)$ : & $v \left[ \left( \lambda'_1 - \lambda'_2 + 2\lambda'_4 - 2\lambda'_5 \right) s_{2\beta} + 2\left( \lambda'_3 + \lambda'_6 + \lambda'_7 \right) c_{2\beta} \right] $ \\
 $\left(H^{\rm S} H^{\rm S} H^{\rm S}\right)$ : & $\left[ \mu_{S1} + 3 \mu_{S2} + v_S \left( \lambda''_1 + 4 \lambda''_2 + 3 \lambda''_3 \right) \right]$ \\
 \hline
 $\left(H^{\rm SM} A^{\rm NSM} A^{\rm NSM}\right)$ : & $v \left[ \left(\lambda_1+\lambda_2\right) s_{2\beta}^2/2 + \left(\lambda_3+\lambda_4\right)\left(1+c_{2\beta}^2\right) - \lambda_5 \left( 2+ s_{2\beta}^2 \right) + \left(\lambda_6 - \lambda_7\right) s_{4\beta} \right]$ \\
 $\left(H^{\rm SM} A^{\rm NSM} A^{\rm S}\right)$ : & $\mathcal{M}_{P,12}^2/v$ \\
 $\left(H^{\rm SM} A^{\rm S} A^{\rm S}\right)$ : & $ 2 v \left[ \left( \lambda'_1 - 2\lambda'_4 \right) s_\beta^2 + \left( \lambda'_2 - 2\lambda'_5 \right) c_\beta^2 + \left( \lambda'_3 - \lambda'_6 - \lambda'_7 \right) s_{2\beta} \right]$ \\
 {\color{blue}$ \left(H^{\rm NSM} A^{\rm NSM} A^{\rm NSM}\right)$} : & {\color{blue} $v s_{2\beta} \left[ \lambda_1 c_\beta^2 - \lambda_2 s_\beta^2 - \left( \lambda_3+\lambda_4+\lambda_5\right) c_{2\beta} + \frac{\lambda_6}{t_\beta} - 2\left(\lambda_6 - \lambda_7\right) s_{2\beta} - \lambda_7 t_\beta \right]$} \\
 {\color{OliveGreen} $\left(H^{\rm NSM} A^{\rm NSM} A^{\rm S}\right)$} : & {\color{OliveGreen} $0$} \\
 $\left(H^{\rm NSM} A^{\rm S} A^{\rm S}\right)$ : & $v \left[ \left( \lambda'_1 - \lambda'_2 - 2 \lambda'_4 + 2\lambda'_5 \right) s_{2\beta} + 2 \left( \lambda'_3 - \lambda'_6 - \lambda'_7 \right) c_{2\beta} \right]$ \\
 {\color{blue} $\left(H^{\rm S} A^{\rm NSM} A^{\rm NSM}\right)$} : & {\color{blue} $2 \left\{ \mu_{11} c_\beta^2 - \left(\mu_{12}+\mu_{21}\right) s_{2\beta}/2 + \mu_{22} s_\beta^2 + \right.$}
 \\ & {\color{blue} $ \quad \left.+ v_S \left[ \left( \lambda'_1 + 2\lambda'_4 \right) c_\beta^2 - \left(\lambda'_3 + \lambda'_6 + \lambda'_7\right) s_{2\beta} + \left( \lambda'_2 + 2\lambda'_5 \right) s_\beta^2 \right] \right\}$} \\
 $\left(H^{\rm S} A^{\rm NSM} A^{\rm S}\right)$ : & $2 v \left( \lambda'_6 - \lambda'_7 \right)$ \\
 $\left(H^{\rm S} A^{\rm S} A^{\rm S}\right)$ : & $ -\left[ \mu_{S1} - \mu_{S2} + v_S \left( \lambda''_1 - \lambda''_3 \right) \right]$ \\
 \hline
 $\left(H^{\rm SM} H^+ H^-\right)$ : & $v \left[ \left( \lambda_1 + \lambda_2 \right) s_{2\beta}^2/2 + \lambda_3 \left(1+c_{2\beta}^2 \right) - \left( \lambda_4+\lambda_5\right) s_{2\beta}^2 + \left(\lambda_6 - \lambda_7\right) s_{4\beta}\right]$ \\
 {\color{blue} $\left(H^{\rm NSM} H^+ H^-\right)$} : & {\color{blue} $v s_{2\beta} \left[ \lambda_1 c_\beta^2 - \lambda_2 s_\beta^2 - \left( \lambda_3+\lambda_4+\lambda_5\right) c_{2\beta} + \frac{\lambda_6}{t_\beta} - 2\left(\lambda_6 - \lambda_7\right) s_{2\beta} - \lambda_7 t_\beta \right] $} \\
 {\color{blue} $\left(H^{\rm S} H^+ H^-\right)$} : & {\color{blue} $2 \left\{ \mu_{11} c_\beta^2 - \left(\mu_{12}+\mu_{21}\right) s_{2\beta}/2 + \mu_{22} s_\beta^2 + \right.$}
 \\ 
 & {\color{blue} $ \quad \left.+ v_S \left[ \left( \lambda'_1 + 2\lambda'_4 \right) c_\beta^2 - \left(\lambda'_3 + \lambda'_6 + \lambda'_7\right) s_{2\beta} + \left( \lambda'_2 + 2\lambda'_5 \right) s_\beta^2 \right] \right\}$} \\
 \hline\hline
 \end{tabular}
 \caption{Trilinear couplings in the extended Higgs basis. Note that the coupling between $\left(H^{\rm NSM} A^{\rm NSM} A^{\rm S}\right)$ is strictly vanishing. The couplings $\left(H^{\rm S} A^{\rm NSM} A^{\rm NSM}\right)$ and $\left( H^{\rm S} H^+ H^-\right)$ have the same value as the coupling $\left(H^{\rm NSM} H^{\rm NSM} H^{\rm S}\right)$. Furthermore, the couplings for $\left(H^{\rm NSM} A^{\rm NSM} A^{\rm NSM}\right)$ and $\left( H^{\rm NSM} H^+ H^-\right)$ have the same value as $1/3$ of the coupling for $\left(H^{\rm NSM} H^{\rm NSM} H^{\rm NSM}\right)$.}
 \label{tab:2HDMStricoup}
 \end{center}
 }
\end{table}

\newpage
\section{Independent quartic couplings} \label{app:2HDMS_quartics}
The quartic couplings are obtained from the scalar potential via
\begin{equation}
 g_{\Phi_i\Phi_j\Phi_k\Phi_l}^2 \equiv \left.\frac{\partial^4 V}{\partial \Phi_i \partial \Phi_j \partial \Phi_k \partial \Phi_l}\right|_{\substack{\Phi_1 = v_1\\\Phi_2 = v_2\\S = v_S}} \;.
\end{equation}
We list the independent quartic couplings which can not be obtained as linear combinations of the entries of the mass matrix or the trilinear couplings in Table~\ref{tab:quartics}.

\begin{table}[h!]
 \begin{center}
 \setlength{\extrarowheight}{4.5pt}
 \begin{tabular}{rl}
 \hline\hline
 $\left(\Phi^i \Phi^j \Phi^k \Phi^l\right)$ : & $\lambda_{\Phi^i \Phi^j \Phi^k \Phi^l}$\\
 \hline
 $\left( H^{\rm NSM} H^{\rm NSM} H^{\rm S} H^{\rm S} \right)$ & $ c_\beta^2 \left[ \lambda'_1 +2 \lambda'_4 - 2 \left( \lambda'_3 + \lambda'_6 + \lambda'_7 \right) t_\beta + \left( \lambda'_2 + 2 \lambda'_5 \right) t_\beta^2 \right]$ \\
 $\left( H^{\rm NSM} H^{\rm NSM} A^{\rm S} A^{\rm S} \right)$ & $ c_\beta^2 \left[ \lambda'_1 - 2 \lambda'_4 - 2 \left( \lambda'_3 - \lambda'_6 - \lambda'_7 \right) t_\beta + \left( \lambda'_2 - 2 \lambda'_5 \right) t_\beta^2 \right]$ \\
 $\left( H^{\rm S} H^{\rm S} A^{\rm S} A^{\rm S} \right)$ & $- \left( \lambda''_1 - \lambda''_3 \right)/2$ \\
 $\left( A^{\rm S} A^{\rm S} A^{\rm S} A^{\rm S} \right)$ & $\left( \lambda''_1 - 4 \lambda''_2 + 3 \lambda''_3 \right)/2$ \\
 \hline \hline
 \end{tabular}
 \caption{Independent quartic couplings in the extended Higgs basis.}
 \label{tab:quartics}
 \end{center}
\end{table}

\section{Mono-$Z$}\label{app:monoZsim}

\begin{figure}[h!]
	\begin{center}
		\includegraphics[width=\linewidth]{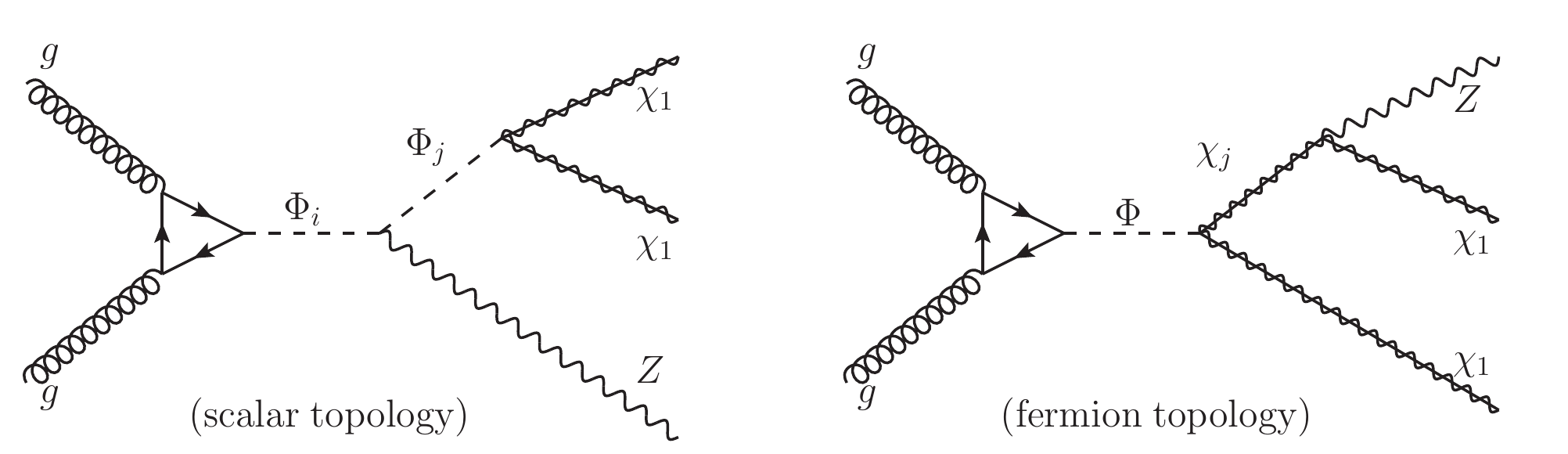}
		\caption{Illustration of resonant decay channels yielding mono-$Z$ signals at the LHC. For the left diagram, $\Phi_i = H$ and $\Phi_j = a$, or $\Phi_i = A$ and $\Phi_j = h$. Decays of the type $A \to Z h_{125}$ are suppressed by alignment, see the discussion in Sec.~\ref{sec:xsec_BR}. This topology is henceforth referred to as the {\it scalar topology}. For the right diagram, $\Phi = H$ or $\Phi = A$, and we refer to this topology as the {\it fermion topology}.}
		\label{fig:feynCascade}
	\end{center}
\end{figure}

In this Appendix, we obtain the future reach in the mono-$Z$ channel in as model independent a fashion as possible. We consider two different topologies giving rise to signals at the LHC, depicted in Fig.~\ref{fig:feynCascade}: The {\it scalar topology} [$gg \to \Phi_i \to Z + (\Phi_j \to \chi_1 \chi_1)$] where the intermediate states are part of the Higgs sector and the {\it fermion topology} [$gg \to \Phi \to \chi_1 + (\chi_j \to \chi_1 + Z)$] where the scalar $\Phi$ in the $s$-channel decays into a pair of new fermions $\chi_1 \chi_j$ and the $\chi_j$ in turn decays into a $Z$-boson and a $\chi_1$ fermion. One may understand the result as derived in a {\it simplified model} framework, where in the case of the scalar topology the SM is extended by two scalars $\Phi_i$, one of them CP-even and one CP-odd, and a new fermion $\chi_1$, and in the case of the fermion topology by one scalar $\Phi$ and two new fermions $\chi_j$ and $\chi_1$. In both cases, $\chi_1$ must be stable on collider time scales and carry no color or electric charge such that it would not deposit energy in the detector, as would naturally be the case if $\chi_1$ is a Dark Matter candidate. It is then straightforward to map the obtained reach to any BSM model containing the required particles by simply computing the corresponding cross section in that model. In particular, in the 2HDM+S the scalar topology arises if one new (stable) fermion $\chi_1$ with mass $m_{\chi_1} \lesssim m_h/2$ ($m_{\chi_1} < m_a/2$) is added and in addition the scalar mass spectrum is arranged such that the primary decay is allowed, i.e. $m_A > m_Z + m_h$ ($m_H > m_Z + m_a$). Likewise, the fermion topology arises if two additional fermions $\chi_j$ and $\chi_1$ are added to the model and the mass spectrum is arranged such that $m_{\chi_j} + m_{\chi_1} < m_\Phi$ and $m_{\chi_1} + m_Z < m_{\chi_j}$, where $\Phi$ can be any of the non SM-like Higgs bosons in the 2HDM+S. 

Before going any further we can already note that, in general, for similar mass spectra, the reach in the scalar topology will be better than in the fermion topology, analogously to what discussed in Ref.~\cite{Baum:2017gbj} for the mono-Higgs signal: In the scalar topology, the \MET\; is given by the transverse momentum of the intermediate scalar $\Phi_j$ produced in the ($\Phi_i \to Z + \Phi_j$) decay back-to-back with the $Z$ boson in the transverse plane. In the fermion topology on the other hand there is no such correlation, it is possible that the $\chi_1$'s in the final state are orientated such that the (vectorial) sum of their respective transverse momenta approximately cancels and hence no \MET\; is observed in the event.

In order to project the reach at the future LHC, we simulate signal samples in both the scalar and the fermion topology using the Monte Carlo generator \texttt{Madgraph5\_aMC@NLO\_v2.5.4}~\cite{Alwall:2014hca} for the generation of the hard event, {\texttt pythia8}~\cite{Sjostrand:2006za,Sjostrand:2014zea} for showering, and the fast detector simulation {\texttt Delphes3}~\cite{deFavereau:2013fsa}. We compare the signal sample for each set of masses with the background taken from Ref.~\cite{ATLAS-CONF-2016-056} in the four different signal categories employed in~\cite{ATLAS-CONF-2016-056}, $\{ (Z \to \mu^+ \mu^-) + \MET, (Z \to e^+ e^-) + \MET \} \otimes \{ {\rm HM}, {\rm LM} \}$, where HM stands for the {\it high mass} and LM for the {\it low mass} search performed in~\cite{ATLAS-CONF-2016-056} by employing the same set of cuts\footnote{Note, that our Monte Carlo configuration differs from that used in Ref.~\cite{ATLAS-CONF-2016-056}; in particular, we use the fast detector simulation {\texttt Delphes3} (with the default ATLAS card in {\texttt Madgraph5\_v2.5.4}) instead of a full detector simulation. This may have numerical impact, although beyond the scope of this work.}.

In the LM search, the final discriminating variable is \MET. In the HM search, the discriminating variable is the transverse mass
\begin{equation}
	\left(m_T^{ZZ}\right)^2 \equiv \left( \sqrt{m_Z^2 + |p_T^{\ell\ell}|^2} + \sqrt{m_Z^2 + |\MET|^2} \right)^2 - \left| \vec{p}_T^{\;\ell\ell} + \vec{E}_T^{\rm miss} \right|^2 \;,
\end{equation} 
where $p_T^{\ell\ell}$ is the transverse momentum of the lepton system. It is worth noting that the LM search employs a lower \MET\; cut, $\MET > 90\,$GeV, than the HM search, $\MET > 120\,$GeV. 

\begin{figure}
	\begin{center}
		\includegraphics[width=\linewidth]{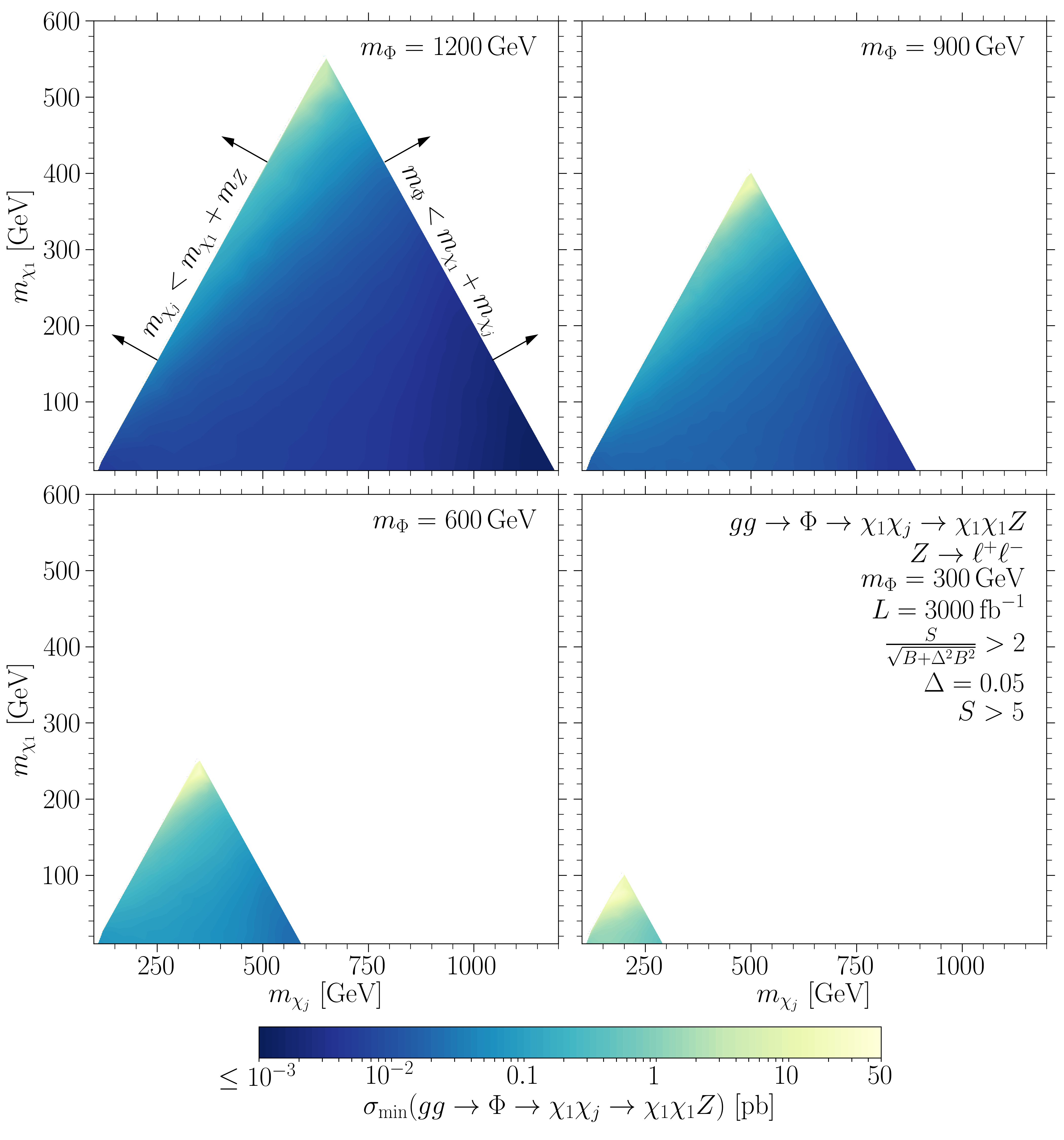}
		\caption{Estimated reach in the fermion topology, cf. Fig.~\ref{fig:feynCascade}, for a luminosity of $L=3000\,{\rm fb}^{-1}$, assuming a systematic uncertainty of the background of $\Delta = 5\,\%$. The reach is shown in the plane of the masses of the involved new fermions $\chi_i$ and $\chi_1$. The different panels correspond to different values of the mass of the scalar $\Phi$ in the $s$-channel as indicated in the respective labels. The color coding shows the best reach for the ($Z + \MET$) final state, i.e. excluding the $Z \to \ell^+\ell^-$ branching ratio. See Appendix~\ref{app:monoZsim} and text for details.}
		\label{fig:reachFermion}
	\end{center}
\end{figure}

We then obtain the future reach at the LHC, $\sigma_{\rm min}$, in both the HM and LM searches by adding the signals in the $Z \to e^+e^-$ and $Z \to \mu^+\mu^-$ categories, comparing to the background taken from Ref.~\cite{ATLAS-CONF-2016-056} rescaled to $L = 3000\,{\rm fb}^{-1}$, and performing a cut-and-count analysis in the signal region where we add an additional lower cut in the discriminating variable of the respective search. We consider a point to be within reach if 
\begin{equation}
	\frac{S}{\sqrt{B + \Delta^2 B^2}} > 2 \quad {\rm and} \quad S > 5 \;,
\end{equation}
where $S$ ($B$) is the number of signal (background) events after all cuts. In the first condition, the two terms in the denominator parameterize the statistical and systematical uncertainty in the background, respectively. In Ref.~\cite{ATLAS-CONF-2016-056}, the systematic uncertainty of the background was approximately $\Delta \approx 8\,\%$. Assuming moderate improvement of the systematic uncertainty of the background, we assume $\Delta = 5\,\%$ for our estimate of the reach.

We show the estimated reach at the LHC for $L = 3000\,{\rm fb}^{-1}$ of data in Fig.~\ref{fig:reachScalar}, located in the main text, for the scalar topology and in Fig.~\ref{fig:reachFermion} for the fermion topology, where for each mass hypothesis we show the better reach out of the HM and LM search. Note, that unless the mass spectrum is close to the kinematic edges indicated in  Figs.~\ref{fig:reachScalar} and~\ref{fig:reachFermion}, the HM search usually has  better reach since it is optimized for larger \MET.

As discussed in the main text, in the scalar topology shown in Fig.~\ref{fig:reachScalar}, the reach is mainly controlled by the mass splitting $\Delta m_\Phi = \left[ m_{\Phi_1} - \left( m_{\Phi_j} + m_Z \right) \right]$, which controls the \MET\; in the process since the ($\Phi_j \to \chi_1 \chi_1$) system and the $Z$ boson are produced back-to-back from the $s$-channel resonance $\Phi_i$. As long as ($2 m_{\chi_1} < m_{\Phi_j}$), such that the ($\Phi_j \to 2 \chi_1$) decay is kinematically allowed, the reach is virtually independent of the mass of $\chi_1$. For $\Delta m_\Phi \gtrsim 300\,$GeV, the reach of the mono-$Z$ search at $L = 3000\,{\rm fb}^{-1}$ is better than $\sigma[gg \to \Phi_i \to Z + (\Phi_j \to \chi_1 \chi_1)] = 10\,$fb.

In the fermion topology, the reach depends on $m_\Phi$, which controls the overall energy scale, as well as the mass splittings at the two vertices, $[m_\Phi - \left( m_{\chi_j} + m_{\chi_1}\right)]$ and $[m_{\chi_j} - \left( m_{\chi_1} + m_Z \right)]$. Since the reach is controlled by three masses, we show it in the $m_{\chi_j}-m_{\chi_1}$ plane for different values of the mass of the parent Higgs boson $\Phi$ in Fig.~\ref{fig:reachFermion}. The reach is significantly weaker than in the scalar topology because the $Z$ boson is produced at the secondary vertex which leads to softer \MET spectra than in the scalar topology where the $Z$ boson is produced back-to-back with the $\Phi_j$ at the primary vertex.

\section{Mapping to the NMSSM}\label{app:NMSSMmap}
The Next-to-Minimal Supersymmetric Standard Model (NMSSM) has the same Higgs sector as the 2HDM+S, albeit its parameters are much more constrained due to supersymmetry. In this Appendix, we provide a mapping of the parameters of the 2HDM+S's scalar potential to those of the NMSSM. 

The scalar potential of the general CP-conserving NMSSM can be written as (cf. Ref.~\cite{Ellwanger:2009dp})
\begin{equation} \begin{split}\label{eq:VNMSSM}
   V_{\rm NMSSM} &= \left| -\lambda H_d \cdot H_u + \kappa S^2 + \mu' S + \xi_F \right|^2 \\
   & \qquad + \left[ m_{H_u}^2 + \mu^2 + \lambda\mu \left( S + {\rm h.c.} \right) + \lambda^2 S^\dagger S \right] H_u^\dagger H_u \\
   & \qquad + \left[ m_{H_d}^2 + \mu^2 + \lambda\mu \left( S + {\rm h.c.} \right) + \lambda^2 S^\dagger S \right] H_d^\dagger H_d \\
   & \qquad + \frac{g_1^2 + g_2^2}{8} \left( H_u^\dagger H_u - H_d^\dagger H_d \right)^2 + \frac{g_2^2}{2} \left| H_d^\dagger H_u \right|^2 + m_S^2 S^\dagger S \\
   & \qquad + \left(\frac{\kappa}{3} A_\kappa S^3 - \lambda A_\lambda S H_d \cdot H_u - m_3^2 H_d \cdot H_u + \frac{m_S'^2}{2} S^2 + \xi_S S + {\rm h.c.} \right).
\end{split} \end{equation}
The dimensionless parameters $\lambda$ and $\kappa$ and the parameter of dimension mass $\mu$ stem from the NMSSM's superpotential. $g_1$ and $g_2$ are the $U(1)_Y$ and $SU(2)_L$ gauge couplings, respectively. The soft SUSY breaking terms are the $m_i^2$ of dimension mass-squared, $\xi_S$ of dimension mass-cubed, and the $A_i$ of dimension mass. The parameter $\xi_F$ of dimension mass-squared and $\mu'$ of dimension mass stem from the $F$-terms.

Note that the much-studied $Z_3$-invariant (or scale-invariant) NMSSM can be obtained by setting $\mu = \mu' = m_3^2 = m_S'^2 = \xi_F = \xi_S = 0$.

In comparison, our parameterization of the 2HDM+S scalar potential, cf. Eqs.~\eqref{eq:V2HDM} and~\eqref{eq:VS}, reads
\begin{equation} \begin{split}
   V_{\rm 2HDM+S} &= m_{11}^2 \Phi_1^\dagger \Phi_1 + m_{22}^2 \Phi_2^\dagger \Phi_2 - \left( m_{12}^2 \Phi_1^\dagger \Phi_2 + {\rm h.c.} \right) \\
   & \qquad+ \frac{\lambda_1}{2} \left( \Phi_1^\dagger \Phi_1 \right)^2 + \frac{\lambda_2}{2} \left( \Phi_2^\dagger \Phi_2 \right)^2 + \lambda_3 \left( \Phi_1^\dagger \Phi_1 \right) \left( \Phi_2^\dagger \Phi_2 \right) + \lambda_4 \left( \Phi_1^\dagger \Phi_2 \right) \left( \Phi_2^\dagger \Phi_1 \right) \\
   & \qquad+ \left[ \frac{\lambda_5}{2} \left( \Phi_1^\dagger \Phi_2 \right)^2 + \lambda_6 \left( \Phi_1^\dagger \Phi_1 \right) \left( \Phi_1^\dagger \Phi_2 \right) + \lambda_7 \left( \Phi_2^\dagger \Phi_2 \right) \left( \Phi_1^\dagger \Phi_2 \right) + \rm{h.c.} \right] \\
   & \qquad + \left( \xi S + {\rm h.c.} \right) + m_S^2 S^\dagger S + \left( \frac{m_S'^2}{2} S^2 + {\rm h.c.} \right) \\
   & \qquad+ \left( \frac{\mu_{S1}}{6} S^3 + {\rm h.c.} \right) + \left( \frac{\mu_{S2}}{2} S S^\dagger S + {\rm h.c.} \right) \\
   & \qquad+ \left( \frac{\lambda''_1}{24} S^4 + {\rm h.c.} \right) + \left( \frac{\lambda''_2}{6} S^2 S^\dagger S + {\rm h.c.} \right) + \frac{\lambda''_3}{4} \left( S^\dagger S \right)^2 \\
   & \qquad+ \left[ S \left( \mu_{11} \Phi_1^\dagger \Phi_1 + \mu_{22} \Phi_2^\dagger \Phi_2 + \mu_{12} \Phi_1^\dagger \Phi_2 + \mu_{21} \Phi_2^\dagger \Phi_1 \right) + {\rm h.c.} \right] \\
   & \qquad+ S^\dagger S \left[ \lambda'_1 \Phi_1^\dagger \Phi_1 + \lambda'_2 \Phi_2^\dagger \Phi_2 + \left( \lambda'_3 \Phi_1^\dagger \Phi_2 + {\rm h.c.} \right) \right] \\ 
   & \qquad+ \left[ S^2 \left( \lambda'_4 \Phi_1^\dagger \Phi_1 + \lambda'_5 \Phi_2^\dagger \Phi_2 + \lambda'_6 \Phi_1^\dagger \Phi_2 + \lambda'_7 \Phi_2^\dagger \Phi_1 \right) + {\rm h.c.} \right].
\end{split} \end{equation}
Identifying the fields
\begin{equation}
   \Phi_1 \to H_u\;, \qquad \Phi_2 \to -i \sigma_2 H_d^*\;, \qquad S \to S\;,
\end{equation}
and globally shifting the scalar potential
\begin{equation}
   V_{\rm 2HDM+S} \to V_{\rm 2HDM+S} + M^4 \;,
\end{equation}
where $M^4$ is a new parameter of dimension mass$^4$, which has no effect on the model's phenomenology, we can map the parameters of the 2HDM+S onto the general NMSSM as given in Tab.~\ref{tab:NMSSMmap}. Note that the mapping requires use of the identity
\begin{equation}
   \left| H_d^\dagger H_u \right|^2 = \left(H_d^\dagger H_d\right) \left( H_u^\dagger H_u \right) - \left| H_d \cdot H_u \right|^2 \;.
\end{equation}

\begin{table}[h!]
   \begin{center}
   \setlength{\extrarowheight}{4.5pt}
   \begin{tabular}{c|ccccccc}
      \hline\hline
      2HDM+S & $m_{11}^2$ & $m_{22}^2$ & $m_{12}^2$ & $M^4$ \\
      NMSSM & $m_{H_u}^2 + \mu^2$ & $m_{H_d}^2 + \mu^2$ & $m_3^2 - \lambda\xi_F$ & $\xi_F^2$ \\
      \hline
     \hline
      2HDM+S & $\lambda_1$ & $\lambda_2$ & $\lambda_3$ & $\lambda_4$ & $\lambda_5$ & $\lambda_6$ & $\lambda_7$ \\
      NMSSM & $\frac{g_1^2 + g_2^2}{4}$ & $\frac{g_1^2 + g_2^2}{4}$ & $-\frac{g_1^2 - g_2^2}{4}$ & $\lambda^2 - \frac{g_2^2}{2}$ & $0$ & $0$ & $0$\\
      \hline
     \hline
      2HDM+S & $\xi$ & $m_S^2$ & $m_S'^2$ \\
      NMSSM & $\xi_S + \xi_F \mu'$ & $m_S^2 + \mu'^2$ & $m_S'^2 + 2 \kappa \xi_F$ \\
      \hline
     \hline
      2HDM+S & $\mu_{S1}$ & $\mu_{S2}$ & $\mu_{11}$ & $\mu_{22}$ & $\mu_{12}$ & $\mu_{21}$ \\
      NMSSM & $2\kappa A_\kappa$ & $2\kappa \mu'$ & $\lambda \mu$ & $\lambda \mu$ & $-\lambda \mu'$ & $-\lambda A_\lambda$ \\
      \hline
     \hline
      2HDM+S & $\lambda_1''$ & $\lambda_2''$ & $\lambda_3''$  \\
      NMSSM & $0$ & $0$ & $4\kappa^2$ \\
      \hline
     \hline
      2HDM+S & $\lambda_1'$ & $\lambda_2'$ & $\lambda_3'$ & $\lambda_4'$ & $\lambda_5'$ & $\lambda_6'$ & $\lambda_7'$\\
      NMSSM & $\lambda^2$ & $\lambda^2$ & $0$ & $0$ & $0$ & $-\kappa\lambda$ & $0$ \\
      \hline\hline
   \end{tabular}
   \caption{Mapping of the 2HDM+S parameters to the general NMSSM. The $Z_3$ NMSSM is obtained by setting $\mu = \mu' = m_3^2 = m_S'^2 = \xi_F = \xi_S = 0$.}
   \label{tab:NMSSMmap}
   \end{center}
\end{table}

\bibliographystyle{JHEP.bst}
\bibliography{nmssmbib}

\end{document}